\documentclass[aps,prb,reprint,superscriptaddress, longbibliography]{revtex4-1}
\usepackage{H1}
\usepackage[pdftex,colorlinks=true]{hyperref}
\hypersetup{citecolor = blue}
\usepackage[normalem]{ulem}
\usepackage{color}
\usepackage[usenames,dvipsnames,svgnames,table]{xcolor}
\usepackage{mathtools}
\usepackage{enumerate}

\newcommand{\Tr}{ \mbox{Tr}}
\newcommand{\vb}{v_B}
\newcommand{\I}{\mathbb{I}}
\newcommand{\ip}{i+1}
\newcommand{\mc}[1]{ { \mathcal {{#1}}}}
\newcommand{\Sz}{S_z^{\rm tot}}

\begin{document}
\title{Operator spreading and the emergence of dissipative hydrodynamics under unitary evolution with conservation laws}

\author{Vedika Khemani}
\affiliation{\mbox{Department of Physics, Harvard University, Cambridge, MA 02138, USA}}
\author{Ashvin Vishwanath}
\affiliation{\mbox{Department of Physics, Harvard University, Cambridge, MA 02138, USA}}
\author{David A. Huse}
\affiliation{\mbox{Department of Physics, Princeton University, Princeton, NJ 08544, USA}}

\begin{abstract}

We study the scrambling of local quantum information in chaotic many-body
systems in the presence of a locally conserved quantity like charge or energy
that moves diffusively. The interplay between conservation laws and scrambling
sheds light on the mechanism by which unitary quantum dynamics, which is
reversible, gives rise to diffusive hydrodynamics, which is a slow dissipative
process. We obtain our results in a random quantum circuit model that is
constrained to have a conservation law. We find that a generic spreading
operator consists of two parts: (i) a conserved part which comprises the weight
of the spreading operator on the local conserved densities, whose dynamics is
described by diffusive charge spreading. This conserved part also acts as a
source that steadily emits a flux of (ii) non-conserved operators. This
emission leads to dissipation in the operator hydrodynamics, with the
dissipative process being the slow conversion of operator weight from local
conserved operators to nonconserved, at a rate set by the local diffusion
current. The emitted nonconserved parts then spread ballistically at a
butterfly speed, thus becoming highly nonlocal and hence essentially
non-observable, thereby acting as the ``reservoir'' that facilitates the
dissipation. In addition, we find that the nonconserved component develops a
power law tail behind its leading ballistic front due to the slow dynamics of
the conserved components. This implies that the out-of-time-order commutator
(OTOC) between two initially separated operators grows sharply upon the arrival
of the ballistic front but, in contrast to systems with no conservation laws,
it develops a diffusive tail and approaches its asymptotic late-time value only
as a power of time instead of exponentially. We also derive these results
within an effective hydrodynamic description which contains multiple coupled
diffusion equations.

\end{abstract}

\maketitle

\section{Introduction}
The nature of quantum dynamics and thermalization in isolated many-body systems is a topic of fundamental interest. Over the last few years, a remarkable confluence of theoretical progress and experimental advances in engineering and controlling isolated many-body quantum systems, especially in cold-atomic gases, has led us to re-examine our understanding of the very foundations of quantum statistical mechanics\cite{Deutsch,Srednicki, Rigol}. On the theory side, research in the field of many-body localization (MBL)\cite{Anderson58, Basko06,  PalHuse,OganesyanHuse, Znidaric,Imbrie2016} has revealed the existence of classes of generically interacting systems that do not obey quantum statistical mechanics, and understanding the nature of different MBL phases\cite{Huse13, PekkerHilbertGlass} and the transition(s) between MBL and thermalizing phases\cite{PalHuse, Kjall14, VHA,PVP, de2016stability,ClarkBimodal,KhemaniCP,KhemaniCPQP} is an active area of research. Complementarily, the development of tools like the AdS/CFT duality\cite{Maldacena_holo, Witten_holo} has led to new perspectives on the dynamics of thermalizing strongly-interacting systems. This duality has been used to relate the physics of information scrambling in black-holes to the process of thermalization in condensed-matter systems of interacting spins and/or particles\cite{HaydenPreskill,SekinoSusskind,HosurYoshida,ShenkerStanfordButterfly,Lashkari,LocalizedShocks,CotlerRM, RobertsStanford,KitaevSYK, SachdevSYK}. 

One lens on the dynamics of isolated many-body quantum systems is provided by studying the spreading of initially-local operators
under the system's unitary time-evolution.   In the Heisenberg picture, an initially local operator $O_0$ evolves into $O_0(t) = U^\dagger(t) O_0 U(t)$ with support on a spatial region that grows with time.  This spreading of $O_0(t)$ is reflected in the growth of the commutator between $O_0(t)$ and a typical local operator $W_x$ with support near position $x$.  If $x$ is well away from the origin, then $W_x$ initially commutes with $O_0$.  We define the ``OTOC'' as the expectation value of the  ``out-of-time-order'' {\it commutator}~\cite{Larkinotoc} 
\begin{equation}
\mathcal{C}(x,t) = \frac{1}{2} \mbox{Tr} \;\rho^{\rm eq}\; [O_0(t), W_x]^\dagger [O_0(t), W_x],
\label{eq:otoc}
\end{equation} 
in an appropriate equilibrium Gibbs state $\rho^{\rm eq}$.  We will almost exclusively consider the infinite temperature ensemble in which case the OTOC reduces to the squared Hilbert-Schmidt norm of the commutator. The OTOC is related to the commutator norm that appears in Lieb-Robinson bounds\cite{Lieb72}, and has received a great deal of attention recently as a diagnostic of information ``scrambling'' in quantum chaotic systems\cite{HaydenPreskill,SekinoSusskind,FawziScrambling,HosurYoshida,ShenkerStanfordButterfly,Lashkari,LocalizedShocks,CotlerRM, RobertsStanford,DoraMoessner, ProsenWeakChaos,KitaevSYK, SachdevSYK}. For a spin-1/2 chain of length $L$, a complete orthonormal basis for all operators is given by the $4^L$ ``Pauli strings'' $\mathcal{S}$, which are products of Pauli matrices on distinct sites.
We can then express our spreading operator in this basis of Pauli strings:
\begin{equation}
O_0(t)=\sum_\mc{S} a_\mc{S}(t)\mc{S} ~.
\label{eq:strings}
\end{equation}
The initially local operator $O_0$ consists only of strings $\mc{S}$ that are the local identity at all sites except one or a few sites near the origin.  But, with time, the strings that dominate this sum grow in spatial extent, containing non-identity local operators at sites out to a ``front'' at a distance from the origin that grows with time.  
The OTOC remains near zero as long as the strings that dominate in $O_0(t)$ contain only local identities near position $x$, but it becomes nonzero once the operator front reaches this position. 

In this paper, we are interested in understanding operator spreading and scrambling in the physically ubiquitous setting of thermalizing systems with one or a few local conservation laws that result in diffusive transport.  This includes, for example, certain Hamiltonian models, since they do conserve energy, and in many cases energy transport is diffusive.  It also includes certain Floquet systems which don't conserve energy but do conserve a charge or spin.  We first investigate this question in a random tensor network spin chain model where the tensors are constrained to obey a single local $U(1)$ spin conservation law. We can derive a number of exact results in this setting, which we conjecture (and numerically verify) should also universally apply to Hamiltonian and Floquet spin chains with diffusing conserved quantities. 

A set of recent papers has studied entanglement and operator dynamics in random tensor network models with no conservation laws\cite{FawziScrambling,AdamCircuit1,opspreadAdam, opspreadCurt}. A subset of these\cite{opspreadAdam, opspreadCurt} showed that, despite the absence of traditional hydrodynamic conservation laws, unitarity alone amounts to a type of conservation law since the operator norm, $\Tr[ O_0^\dagger(t) O_0(t)]$, is conserved which means that the total weight of the operator on all Pauli strings, $\sum_\mc{S} |a_\mc{S}|^2 $,  is conserved. This leads to an emergent ``hydrodynamical'' picture for describing operator spreading wherein, in one dimension, the dynamics of the operator front can be described by a distribution of biased random walkers (where the bias reflects the fact that it is more likely for an operator string to grow rather than to shrink)\cite{opspreadAdam, opspreadCurt}, while in higher dimensions the front is modeled by a random growth model\cite{AdamCircuit1,opspreadAdam}.  This leads to a picture in which the operator front propagates ballistically with ``butterfly'' speed $v_B$, while in low dimensions the width of the front grows as a sublinear power of time ($\sim\sqrt{t}$ in 1D)\cite{opspreadAdam, opspreadCurt}.  This means that the bulk of the total weight of $O_0(t)$ is contained in the spatial region lying within the ``light cone''  defined by the front whose spatial linear size grows linearly with time.
Further,  $O_0(t)$ is scrambled within the spatial region defined by the light cone and, with respect to some (but not all) measures, resembles an unconstrained random operator in this region\footnote{Differences between $O_0(t)$ and a random operator are apparent when considering the spectrum of $O_0(t)$, which is highly degenerate and always the same as the spectrum of $O_0$ due to the unitary time-evolution.  Also, the operator entanglement, although a ``volume law'', is only a fraction of that of a random operator\cite{jonay}.}. 

In this work, we build on the picture above and study the interplay between the ``actual'' hydrodynamics governing the dissipative diffusion of conserved charges and the ``emergent'' hydrodynamics of unitary operator spreading in systems with conservation laws.  Starting from a local operator that contains the conserved charge, we find that there is again a ballistically propagating front describing the operator spreading.  However, unlike the fully nonconserved case, $O_0(t)$ has a significant (decreasing slowly as a power law in time) amount of weight on operator strings whose fronts lag far behind the main operator front. This can be understood as follows:  Since conserved charges spread diffusively, the parts of $O_0(t)$ that overlap with the conserved charges are ``left behind'' in a region of linear size $\sim\sqrt{t}$ near the origin, even while the main operator front has reached distance $v_Bt$.  But the total weight of $O_0(t)$ on these local conserved operators decreases as a power-law in time $\sim t^{-d/2}$ in $d$ dimensions as the conserved density spreads diffusively.  This loss of operator weight makes this diffusion effectively nonunitary and thus dissipative.  But the full dynamics of the system is unitary so the full operator weight is not lost.  Instead the dissipation due to the diffusive currents of the conserved density steadily converts operator weight from conserved to non-conserved operators, thus emitting
a ``flux'' of non-conserved operators that then spread ballistically.  This emitted flux is proportional to the square of the diffusive current, as expected for ``Ohmic'' dissipation.  One point of view is that once the non-conserved operators start spreading ballistically, they rapidly become so highly nonlocal that they stop being observables.  Then they can be viewed as effectively random, functioning as the bath whose entropy increases due to the dissipative diffusion. 

An important point is that the conversion of operator weight from locally conserved to non-conserved (and non-local) happens at a power-law \emph{slow} rate set by the diffusion current, and thus the dissipation shows up as a slow \emph{hydrodynamic} process in the operator dynamics. By contrast, in an unconstrained model, the dissipative conversion of operator weight from local to non-local happens on an $O(1)$ time-scale and thus does not show up as a slow hydrodynamic process. An explicit elucidation of how unitary time evolution with conservation laws can lead to \emph{slow} dissipative hydrodynamics is one of the main contributions of this work.   

Moreover, in the models we study here, we can also study the unitary ``inner workings'' of this effective bath, following the spreading of the non-conserved parts of the operator.
The bulk of these non-conserved operators are emitted almost immediately, near time $t=0$, and these grow to form the leading operator front at distance $\vb t$ at time $t$.  However, as a result of the conservation law, there is still significant weight left on the conserved parts even after the initial emission, and these continue to emit non-conserved operators at a slow rate that  decreases only as a power law in time. Thus, those non-conserved parts of the operator that are emitted at a later time $t_e$ have a front that lags behind the main operator front by distance $v_Bt_e$.  This leads to a power-law tail in the operator profile behind the front.
By contrast, the unconstrained random circuit model does not show such a power-law tail, since in that case there are no slow dissipative modes so no significant part of the operator has a front that lags substantially behind the main operator front.
Fig.~\ref{fig:opshape} shows a sketch of the operator profile depicting all three regimes: (i) the diffusive  conserved charges remaining near the origin, (ii) the ballistically moving front, and (iii) the power law tail behind the front.  

The picture that emerges from our study is of multiple coupled hydrodynamic equations:  The first is the dissipative diffusion of the conserved quantity.  This dissipation serves as the source of the ballistically spreading nonconserved operators.  The dynamics of the fronts of these nonconserved operators is biased diffusion in 1D and a random growth model in higher dimensions\cite{AdamCircuit1, opspreadAdam}.  Moreover, we find that the local operator content in the \emph{interior} of the spreading operator is also governed by two coupled noisy diffusion equations, with the leading front of the operator serving as a moving boundary condition on these equations.  And finally, there is at least one more ``layer'' of this hydrodynamics that governs the entanglement dynamics of the spreading operator\cite{jonay}.

The spatial profile of the spreading operator described above is also reflected in the behavior of the OTOC $\mathcal{C}(x,t)$ defined in Eq~\eqref{eq:otoc}. We show that $\mathcal{C}(x,t)$ increases sharply when the ballistic operator front reaches $x$ but, as a result of the power-law tails behind the main operator front, it only approaches its asymptotic late-time value as a power-law in time. This is contrast to systems without conservation laws where there are no such power-law diffusive tails\cite{opspreadAdam, opspreadCurt}.  
Our results help explain the numerical observations in Ref.~\onlinecite{FradkinHuse} where this late-time power-law in the OTOC was observed in systems with conservation laws.  

We note that much recent work has focused on computing the OTOC at low temperatures in systems with energy conservation, and  bounds on the growth of the OTOC have been derived in this setting\cite{chaosbound}. On the other hand, random circuit models do not conserve energy and thus the ``infinite temperature'' Gibbs ensemble is the only meaningful one for such models. Nevertheless, for circuit models endowed with extra conservation laws like total spin/charge, the dynamics can be resolved into different spin sectors with the ``chemical potential'' $\mu$ now playing a role somewhat analogous to the inverse temperature $\beta$. We mostly focus on $\mu=0$ in this work, while briefly addressing the $\mu \neq 0$ case.  

We note that several papers have recently noted that quantum information can spread ballistically in systems with diffusively relaxing conserved charges\cite{KimHuse,  KnapScrambling, LuitzScrambling, SwingleChowdhury, PatelDiffusiveMetal,AleinerOTOC}. This, by itself, is not that surprising since even MBL systems with no charge transport can show logarithmic spreading of entanglement\cite{BardarsonPollmannMoore,Znidaric}.  
Of these papers, Refs.~\onlinecite{SwingleChowdhury, PatelDiffusiveMetal,AleinerOTOC} study a weakly interacting diffusive metal and, using a perturbative semiclassical scattering calculation, relate the butterfly speed $\vb$ to the diffusion constant of the metal via a ``Lyapunov  exponent" which characterizes the exponential growth of ``chaos'' in this semiclassical setting as measured by the growth of the OTOC before the front arrives: $C(x,t) \sim e^{-\lambda_L (x-\vb t)}$. 

On the other hand, Refs.~\onlinecite{KimHuse,  KnapScrambling, LuitzScrambling} numerically study fully quantum spin-chains which cannot be treated semiclassically, and for which no prolonged period of exponential growth in the OTOC has been observed to date (the existence of a fully quantum Lyapunov exponent in spatially extended systems with small local Hilbert spaces and only short-range interactions, while perhaps expected, is a presently unresolved question\cite{ProsenWeakChaos,Khemani_VDLE,Swingle_OpSpreadMPO}). In this fully quantum setting where semiclassical analytical methods don't apply, the aforementioned numerical papers\cite{KimHuse,  KnapScrambling, LuitzScrambling} generally treat the diffusive charge relaxation and the ballistic information spreading as two independent numerical observations that do not interface with one another.  

One of our main contributions in this work is to connect the diffusive charge dynamics with the ballistic front spreading in strongly quantum systems with local conservation laws into a composite picture for the operator profile, showing how the different regimes connect at different time and length scales.  
\emph{En route}, we elucidate how reversible unitary dynamics can still display dissipative hydrodynamic modes, wherein the dissipative process is the conversion of operator weight from locally observable conserved parts to non-local and essentially unobservable non-conserved parts at a slow rate set by the local diffusion current of the conserved densities. Thus while the von Neumann entropy of the full system is conserved under the unitary dynamics, we show how the local ``observable'' entropy can still increase at a slow hydrodynamic rate,  very concretely illustrating a resolution of the fundamental tension between unitarity and dissipation in closed quantum systems.    

The balance of this paper is structured as follows. We begin in Section~\ref{sec:NetworkModel} with a description of our constrained random unitary circuit model which has a local $U(1)$ conservation law total for total spin. This is followed by a detailed discussion of the spatial profile of spreading operators in Section~\ref{sec:opshape}. We show that the conserved charges evolve diffusively under the action of the circuit.  The coupling between the charge conservation and unitarity produces a steady flux of operator weight from the diffusively spreading conserved components to ballistically spreading nonconserved components, leading to a power law tail in the spatial profile of the operator weight. We present the coupled hydrodynamics describing this process. Next, in Section~\ref{sec:raisingandspin}, we expose another layer of structure in the spreading operator by studying the local operator content of the highly nonlocal operator strings \emph{within} the light cone defined by the leading ballistic front. We find that the distribution of different local operators is itself governed by a set of coupled noisy diffusion equations that ``turn on'' once the front passes a given position. In Section~\ref{sec:otoc}, we turn to a discussion of OTOC's in this model, and find that the diffusive processes governing the shape and internal structure of the spreading operators lead to a late time diffusive tail in  OTOC's involving the conserved charges.  We numerically verify that the universal aspects of our results also apply to more ``physical'' spin chains with energy/charge conservation in Section~\ref{sec:physical}, and conclude in Section~\ref{sec:conclusion}.  Some additional details are discussed in the Appendices.

\section{ Random unitary circuit Model}
\label{sec:NetworkModel}
As an explicit example where we can obtain analytical results, we consider a one-dimensional chain of length $L$ sites where the degree of freedom on each site is the direct product of a spin-1/2 (or qubit) and a qudit with Hilbert space dimension $q$.  
The time-evolution is constrained to conserve the total $z$ component of the spin-1/2's, which we call $S_z^{\rm tot}$, while the qudits are not subject to any conservation laws.  We note that we include the qudit at each site because some results can only be obtained analytically in the large-$q$ limit, although some of our results do apply for all $q$, including the case $q=1$ that has no qudits. For definiteness we take a finite $L$, but we will be interested in the behavior in the limit of infinite $L$.  

\begin{figure}
  \includegraphics[width=\columnwidth]{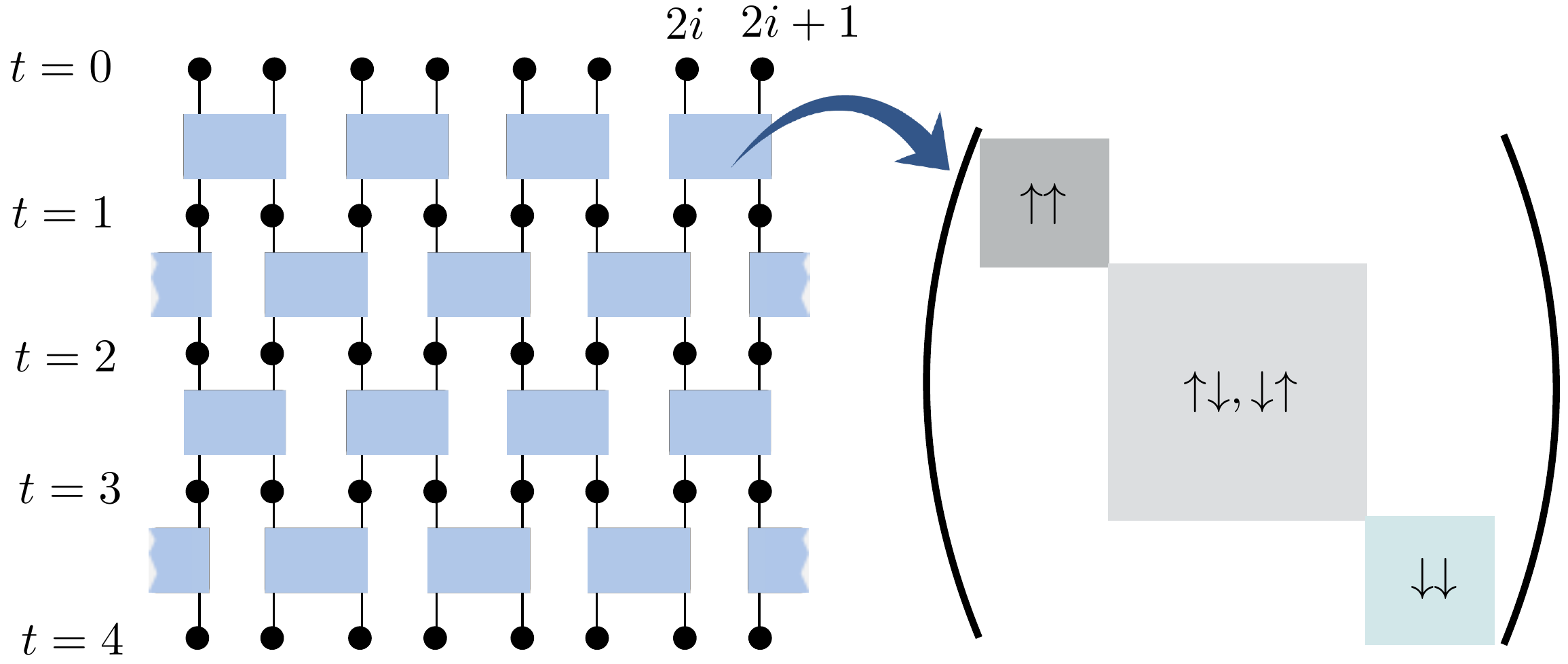}
  \caption{\label{fig:circuit} 
Left: a diagram of the random unitary circuit.  Each site (black dot) is the direct product of a two-state qubit and a $q$-state qudit.  Each gate (blue box) locally conserves $S_z^{\rm tot}$, the total $z$ component of the two qubits it acts upon, and is thus a block-diagonal unitary of the form shown on the right, with each block of each gate independently Haar-random.  The smaller blocks do not flip the qubits and thus operate only on the two qudits, while the larger block also produces $S_z^{\rm tot}$-conserving qubit ``flip-flops''. }
\end{figure}

Generalizing from Refs.~\onlinecite{FawziScrambling,AdamCircuit1,opspreadAdam, opspreadCurt}, the time-evolution is governed by a random quantum circuit comprising staggered layers of two-site unitary gates acting on even and odd spatial bonds at even and odd times respectively (see Fig.~\ref{fig:circuit}). The time evolution operator is given by $U(t) = \prod_{t'= 1}^{t} U(t', t'-1)$, where 
\begin{equation}
U(t', t'-1) = \begin{cases}
  \prod_{i}  U_{2i,2i+1}   & \text{if $t'$ is even},\\
  \prod_{i}  U_{2i-1,2i}   & \text{if $t'$ is odd}.
  \end{cases}
\end{equation}
As a result of the conservation law, each two-site unitary gate $U_{i, i+1}$ is a  $(4q^2 \times 4q^2)$-dimensional block-diagonal matrix.  Labeling the spin state on each site $i$ as $(\uparrow a)_i$ or $(\downarrow b)_i$, where the first label is the spin state in the Pauli $z$ basis and the second label is the qudit state, the structure of $U_{i, i+1}$ looks like: (i) a $(q^2\times q^2)$ block acting in the $(\uparrow a)_i\otimes (\uparrow b)_{\ip}$ subspace, (ii) a $(2q^2\times 2q^2)$ block acting in the $(\uparrow a)_i\otimes (\downarrow b)_{\ip},(\downarrow a)_i\otimes (\uparrow b)_{\ip}$ subspace, and (iii) a $(q^2\times q^2)$ block acting in the $(\downarrow a)_i\otimes (\downarrow b)_{\ip}$ subspace.  Each of these blocks is a Haar-random unitary, and each block in each two-site gate is chosen independently of all others. 

To characterize the time-evolution of local operators, it is useful to define a complete orthonormal basis of operators on each site.  For the spin, we can use the Pauli matrices on each site to define an onsite basis as $$\{\sigma_i^{\mu=0,1,2,3}\} \equiv \{\I_i, r_i, l_i,z_i\}=\{ \mathbb{I}_i,\frac{\sigma_i^+}{\sqrt{2}}, \frac{\sigma_i^-}{\sqrt{2}}, \sigma_i^z\},$$ so $r_i$ and $l_i$ are suitably normalized spin raising and lowering operators, respectively.  These basis operators all have a definite $\Delta S_z^{\rm tot}$ (such as ``raise/lower by one'') under the $U(1)$ symmetry that conserves $S_z^{\rm tot}$ and are thus more convenient than the Pauli $\sigma_i^{x/y}$ matrices for characterizing the $U(1)$-conserving dynamics.  For the qudit, one can construct higher-dimensional generalizations of the Pauli matrices $\{\Sigma_i^{\mu = 0, 1, \cdots q^2-1}\}$ that are normalized such that $\Tr(\Sigma_i^{\mu\dagger} \Sigma_i^\nu)/q=\delta_{\mu\nu}$. Then, the tensor product $B_i^{\mu \nu} \equiv \sigma_i^\mu \otimes \Sigma_i^\nu$ generates a local basis for the $4q^2$ operators acting on each composite site $i$, denoted in shorthand as $(\I \Sigma^\nu)_i$,  $(r \Sigma^\nu)_i$, $(l \Sigma^\nu)_i$ and $(z \Sigma^\nu)_i$. Using this basis, the time evolved operator $O(t)$ can be expanded as:
\begin{align}
O(t) = \sum_\mathcal{S} a_\mathcal{S} (t) \mathcal{S},
\label{eq:Otexpand}
\end{align}
where each generalized Pauli string $\mathcal{S}$ is  one of $(4q^2)^L$ basis product operators, $\prod_iB_i^{\mu_i \nu_i}$.  Since the basis strings satisfy $\Tr [\mc{S}^\dagger \mc{S}']/(2q)^L = \delta_{\mc{S} \mc{S}'}$, the coefficients $a_\mathcal{S}$ can be obtained as $a_\mathcal{S}(t) = \Tr[ \mc{S}^\dagger O(t)]/(2q)^L$. Finally,  we normalize the initial operator $O_0$ such that $\Tr [O_0^\dagger O_0] = (2q)^{L}$ which, by the unitarity of the dynamics, implies that the total weight of $O(t)$ on all strings $\mc{S}$ is also normalized to 1: 
\begin{equation}
\sum_\mathcal{S} |a_\mc{S}(t)|^2 = 1.
\label{eq:onorm}
\end{equation}
This sum rule is the effective conservation law due to unitarity\cite{opspreadAdam,opspreadCurt}.

There are a few classes of operators on site $i$ that evolve differently under the action of this conserving  unitary circuit. First, $(z \I)_i$ measures the local conserved charge, and $(\I \I)_i$ is the identity operator. The conservation law implies that $S_z^{\rm tot}$ is conserved so that
\begin{equation}
S_z^{\rm tot}= \sum_i (z \I)_i, \qquad U(t)^\dagger \Sz U(t)  = \Sz,
\label{eq:Szcons}
\end{equation}
and the operators  $(\I\I)_i(\I\I)_{i+1}$, $(z \I)_i(z \I)_{i+1}$,  and $[(z\I)_i(\I\I)_{i+1}~+~(\I\I)_i(z\I)_{i+1}]/\sqrt{2}$  are left invariant by the action of all local gates $U_{i, \ip}$.  Further, if one starts with an operator with a definite $\Delta S_z^{\rm tot}$ under the U(1) symmetry  (for example, $(r \Sigma^\nu)_i$ raises the spin by one), the action of the circuit preserves this $\Delta S_z^{\rm tot}$.  Appendix~\ref{sec:circuitaction} summarizes the action of $U_{i, i+1}$ on all possible two-site operators. 

It will be subsequently useful to separate the spreading operator into conserved and non-conserved pieces. To intuitively understand this separation, consider an initial density matrix 
\begin{equation}
\rho(0) = (\I_{\rm all} + A{O_0})/{(2q)^L},
\label{eq:rhoeq_pert}
\end{equation} 
where $\I_{\rm all}$ is the background equilibrium state which is the identity on the full system;
$AO_0$ is a traceless local on-site perturbation at the origin to this equilibrium state, with $O_0$ normalized such that $\Tr[O_0^\dagger O_0] = (2q)^L$, and $A$ is the amplitude of this perturbation, which must be small enough so $\rho$ remains non-negative.  The system conserves $S_z^{\rm tot}$ \eqref{eq:Szcons}  so that 
\begin{align}
\langle \Sz \rangle (t) =  A\frac{\Tr [O_0(t) \Sz]}{ (2q)^L} = \langle \Sz \rangle(0)   = A\frac{\Tr[ O_0 \Sz]}{ (2q)^L},
\label{eq:Otcons}
\end{align}
where $\langle \rangle(t)$ denotes expectation values in the state $\rho(t)$.  If the perturbation injects some local charge at the origin then, on general grounds, we expect this ``extra'' charge to spread diffusively so that $\langle  (z\I)_x \rangle \sim \frac{1}{\sqrt{t}}e^{-x^2/4Dt}$. We will see how this diffusion arises in operator language and explore its consequences for operator dynamics in this system.    

To separate the part of $O_0(t)$ that has overlap with the conserved charges,  we define $a^c_i(t)$ as the amplitude of the conserved charge $(z\I)_i$ in the operator expansion of $O_0(t)$:
\begin{equation}
a^c_i(t) \equiv \frac{1}{(2q)^L}\Tr [O_0(t) (z\I)_i] = \langle  (z\I)_i \rangle (t) .
\end{equation}
The conserved charges which act as $(z\I)_i$ on site $i$ and as the identity everywhere else are a subset of the full basis of operator strings, and the ``conserved  part'' of $O_0(t)$ is defined as the part of $O_0(t)$ with weight on these basis strings:
\begin{equation}
O_0^c(t) = \sum_i a_i^c(t) (z\I)_i,
\label{eq:Oc}
\end{equation}
with the ``non-conserved part'' being the rest, $O_0^{\rm nc}(t) = O_0(t) - O_0^c(t)$.  Finally, the conservation law \eqref{eq:Otcons} requires
\begin{align}
\sum_i a^c_i(t) = \mbox{constant}. 
\label{eq:chargecons}
\end{align}
The dynamics of operator spreading in this model is governed by the interplay between (i) this explicit charge conservation \eqref{eq:chargecons} which is a sum rule on the {\it amplitudes} of the conserved operators, and (ii) the conservation of the operator weight (which follows from unitarity)  which is a sum rule on the {\it squares} of the amplitudes of {\it all} basis strings \eqref{eq:onorm}. 

\section{``Shape'' of Spreading Operators}
\label{sec:opshape}
A complete characterization of the spreading and scrambling of an initially local operator $O_0$ requires knowledge of the exponentially-many coefficients $a_\mc{S}(t)$ in the expansion of $O_0(t)$ as in Eq.~\eqref{eq:Otexpand}. Instead of doing this, we first consider a more coarse-grained approach and define the ``right-weight'' $\rho_R(i,t)$ as the total weight in $O(t)$ of basis strings that end at site $i$, which means that they act as the identity on all sites to the right of site $i$, but act as a non-identity on site $i$. This is the weight on all strings of the form  $\mc{S}\sim (\prod_{k<i}B_k^{\mu_k \nu_k})(n)_i (\I\I)_{i+1}  (\I\I)_{i+2} \cdots (\I \I)_L$, where $(n)_i \in \{(r\Sigma^{\mu})_i, (l\Sigma^{\mu})_i, (z\Sigma^{\mu})_i, (\I\Sigma^{\mu>0})_i  \}$ denotes a non-identity operator on site $i$:
\begin{equation}
\rho_R(i,t) = \sum_{\substack{{\text{strings $\mc{S}$ with } }\\ \mathclap{\text{ rightmost non-}}\\ \mathclap{\text{identity on site $i$}}}} |a_\mc{S}|^2, \qquad \,\,\, \sum_i \rho_R(i,t)=1. 
\label{eq:rw}
\end{equation}
The conservation law on $\rho_R(i,t)$ which follows from unitarity \eqref{eq:onorm} gives $\rho_R(i,t)$ the interpretation of an emergent local conserved ``density'', and Refs.\cite{opspreadAdam, opspreadCurt} showed that the (hydro)dynamics for $\rho_R(i,t)$ is governed by a biased diffusion equation. Of course, one can analogously define the left-weight $\rho_L(i,t)$ and, together, these can be used to characterize some illuminating aspects of the spatial structure of the spreading operator (we will consider other measures probing the local operator content inside the spreading operator in the next  section).

\begin{figure}
  \includegraphics[width=\columnwidth]{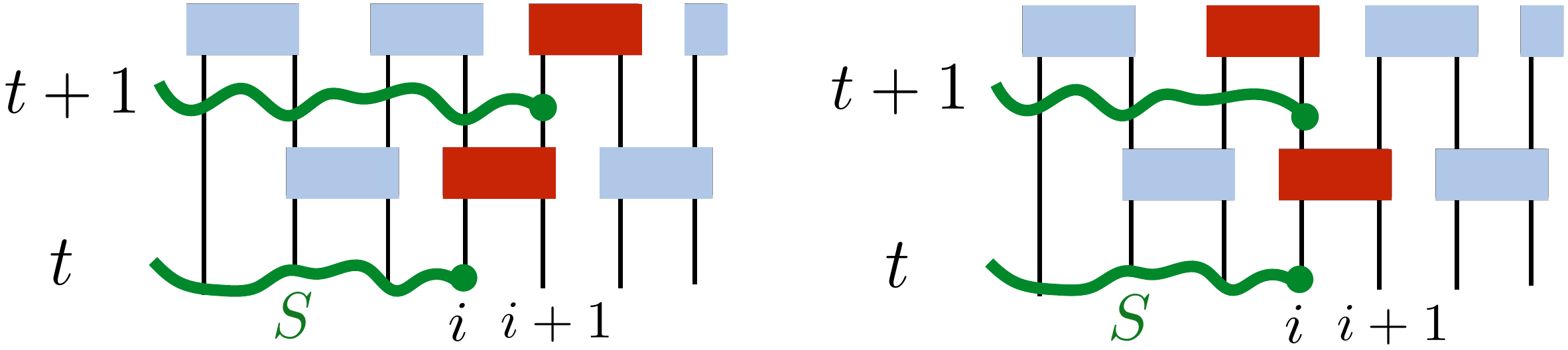}
  \caption{\label{fig:front} A Pauli string $S$ (green) with its rightmost non-identity operator on site $i$ at time $t$ has its right-front on the gate $(i,i+1)$ (colored red). Under the action of the circuit, the front moves forwards if the endpoint of the string moves to $(i+1)$ at time $t+1$ (left), while the front moves backwards if the action of the circuit leaves the endpoint at site $i$ (right). }
\end{figure}

It is instructive to first understand the dynamics of $\rho_R(i,t)$ in an unconstrained random circuit model in 1D with local Hilbert space dimension $q$.  It was shown in Refs.\cite{opspreadAdam, opspreadCurt} that the weighted distribution of the endpoints of the basis strings in the operator time-evolve as biased random walkers, where the bias reflects the fact that it more likely for the strings to grow rather than to shrink.  If we define the ``right-front'' of a string as the location of the rightmost unitary \emph{gate} that sees a non-identity, one can obtain the probabilities for the front to move forwards or backwards by noting that under the action of the front gate, all the $(q^4-1)$ non-identity operators at the gate are produced with equal weights,
on average.  If the front gate is $U_{i, i+1}$,  only $q^2-1$ of the operators that can be produced by this gate act as the identity on the right site of the gate $(i+1)$, and each of these outcomes
results in the front moving a step backwards,  an event with probability $p  = \frac{1}{(q^2+1)}$.  Because we define the front at living on gates (bonds) rather than sites, the even-odd structure of the circuit in time implies that the front moves backwards when the endpoint of a string does not grow (illustrated in Fig.~\ref{fig:front}).  

Then, from the theory of random walks, the mean location of the front after time $t$ is $\langle x \rangle = (1-2p)t \equiv \vb t$, which defines the butterfly speed $\vb$, and the width of the front grows as $\sim\sqrt{D_\rho t}$, where $D_\rho = (1-\vb^2)/2$ is the diffusivity of the resulting biased random walk. Evidently, the front location can be described by the emergent random walk hydrodynamics\cite{opspreadAdam, opspreadCurt} $\partial_t \rho_R(x,t) = \vb \partial_x \rho_R(x,t) + D_\rho \partial_x^2 \rho_R(x,t)$, and the fact that the right side of this equation is a total spatial derivative reflects the conservation of the emergent density $\rho_R$ \eqref{eq:rw}. Fig~\ref{fig:opshape}(c) shows a sketch of the Haar-averaged front for this random circuit evolution which, in the scaling limit $t,x\rightarrow\infty$ for $x \approx v_Bt$ takes the form\cite{opspreadAdam,opspreadCurt}
\begin{equation}
{\rho_R^{\rm rand}(x,t)} = \frac{1}{\sqrt{4\pi D_\rho t}}e^{-\frac{(x-v_Bt)^2}{4D_\rho t}}. 
\label{eq:rwRand}
\end{equation}
In the limit ${q\rightarrow \infty}$, the front becomes sharp and $\vb\sim~1- \frac{2}{q^2} \rightarrow 1$ so that the front deterministically moves forwards for all basis strings at each time step in this limit (Fig~\ref{fig:opshape}(d)).  Note that the geometry of the circuit limits the operator growth to at most one site on the left and right ends per time step.  This imposes a strict upper bound on $\vb$ set by the ``causal limit speed'' $v_{\rm CL}=1$, and $\vb$ goes to this limit as ${q\rightarrow \infty}$. 

Let us now turn to how this picture changes in the presence of an explicit $U(1)$ conservation law with diffusively spreading charges. 

\subsection{Spreading of the local conserved charge}
Let us begin with the case when the initial operator is a conserved $U(1)$ charge located at the origin, $O_0 = (z\I)_0$. In this case, the spatial structure of the operator shows three regimes: (i) a ballistic front, (ii) a power-law tail behind the front, and (iii) diffusively spreading charges near the origin.  

\subsubsection{Diffusion of conserved charge}
We start by discussing the diffusive dynamics of the conserved parts of the operator. For the initial operator $O_0 = (z\I)_0$, the initial conserved amplitudes are  $a^c_i(0) = \delta_{i0}$ and the conservation laws \eqref{eq:Otcons},\eqref{eq:chargecons}  give 
\begin{equation}
\sum_i a^c_i(t)  = 1~.
\label{eq:acons}
\end{equation}
When viewed as a perturbation to an equilibrium state \eqref{eq:rhoeq_pert}, $O_0 = (z\I)_0$ creates an excess of charge at the origin which should spread diffusively. Thus, on general grounds, at late times we expect $\langle (z\I)_x \rangle (t) = a_x^c(t) \sim \frac{1}{\sqrt t } e^{-x^2/(4D_ct)}$, where $D_c$ is the charge diffusivity.  We now see how this arises. 

To understand the evolution of $a_i^c(t)$, consider the action of a particular gate $U_{12}$ on the superposition $[a^c_1 (z\I)_1~+~a^c_{2} (z\I)_2]$.  It can be shown (Appendix~\ref{sec:consamps}) that the action of a gate makes the average amplitudes $\overline{a^c}$ equal on the two sites acted upon by it, while preserving the sum of amplitudes. That is, after the action of the gate, the Haar-averaged amplitudes are 
\begin{equation}
\overline{a^c_1(t+1)} = \overline{a^c_2(t+1)}= \frac{\overline{a_1(t)}+\overline{a_2(t)}}{2}
\label{eq:acircuit}
\end{equation}
and the gate has produced a current between these two sites of the conserved charge that is $\sim (a_1(t)-a_2(t))\sim \partial_x a(x,t)$. Note the ``smoothing'' action of the circuit which locally makes the averaged amplitudes equal, and thus reduces $\partial_x a(x,t)$ in time.    
This action gives a binomial charge distribution (Appendix~\ref{sec:consamps}):
\begin{align}
\overline{a_i^c(t)} = \frac{1}{2^t}   {{t-1}\choose{\floor{\frac{i+t-1}{2}}}},
\label{eq:abinomial}
\end{align}
which is an exact result true for all $q$ including $q=1$.  If we coarse-grain, 
in the scaling limit $x,t\rightarrow \infty$ we get
\begin{equation}
\overline{a^c(x,t)} =\sqrt{ \frac{1}{2\pi t}} e^{-\frac{x^2}{2t}},
\label{eq:adiff}
\end{equation}
showing diffusion of the conserved charge with diffusion constant $D_c = 1/2$.  If we consider similar random circuits acting on a system in higher dimensions $d>1$, a similar diffusive behavior will be present, just with diffusion along all directions. It is noteworthy that as a consequence of the conservation law, the Haar-averaged \emph{amplitudes} $\overline{a_i^c(t)}$ are non-zero \eqref{eq:abinomial}. In an unconstrained random circuit, only the squared \emph{weights} $\overline{|a_S|^2}$ survive Haar averaging while all amplitudes average to zero. Likewise, in our model with $\Sz$ conservation,  certain off-diagonal products of amplitudes $\overline{a_S a_{S'}}$ can also survive Haar averaging (Appendix~\ref{sec:consamps}), while all such averages are zero in the unconstrained model. 

\subsubsection{Ballistic front and power law tails}
We now turn to the effect of the diffusive dynamics on the shape of the spreading operator as measured by $\rho_R(i,t)$. It is important to note that $\rho_R(i)$ measures the \emph{weights}, $|a_\mc{S}|^2$, of \emph{all} operator strings ending on $i$, while the preceding discussion was about the diffusive dynamics of the \emph{amplitudes}, $a_i^c$, of \emph{only} the conserved charges in the expansion of $O_0(t)$; these conserved operators are strings of length one site. 
Of course, the weight of $O_0(t)$ on a conserved charge at site $i$ contributes to $\rho_R(i,t)$ and it is convenient to separate out this contribution:
\begin{align}
{\rho_R(i,t)} &= {|a_i^c(t)|^2} + {\rho_R^{\rm nc}(i,t)} \nonumber \\
&  \equiv {\rho^{\rm c}(i,t)} + {\rho_R^{\rm nc}(i,t)} ,
\label{eq:rhoparsed}
\end{align}
where $\rho_R^{\rm nc}$ denotes the right-weight from all ``non-conserved'' operator strings that are \emph{not} one of the conserved charges $(z\I)_i$. Defining $\rho^c_{\rm tot} \equiv \sum_i {|a_i^c(t)|^2}$ and $\rho^{\rm nc}_{\rm tot} \equiv \sum_i {\rho_R^{\rm nc}(i,t)}$, we know from \eqref{eq:rw} that 
\begin{equation}
\rho^c_{\rm tot} + \rho^{\rm nc}_{\rm tot} = 1. 
\label{eq:rhoparsedcons}
\end{equation}
We will show that the total weight of $O_0(t)$ on the conserved charges, $\rho^c_{\rm tot}$, decreases as a power law in time; these then act a source, continuously emitting a flux of non-conserved operators that spread ballistically and contribute to $\rho_{\rm tot}^{\rm nc}$.\footnote{Note, if the initial operator is Hermitian, the amplitudes $a_i^c(t)$ all remain real, but here we are using absolute value brackets so that the formulae apply even to non-Hermitian operators.  Later we will be considering raising and lowering operators which are indeed non-Hermitian.}

\begin{figure*}
  \includegraphics[width=\textwidth]{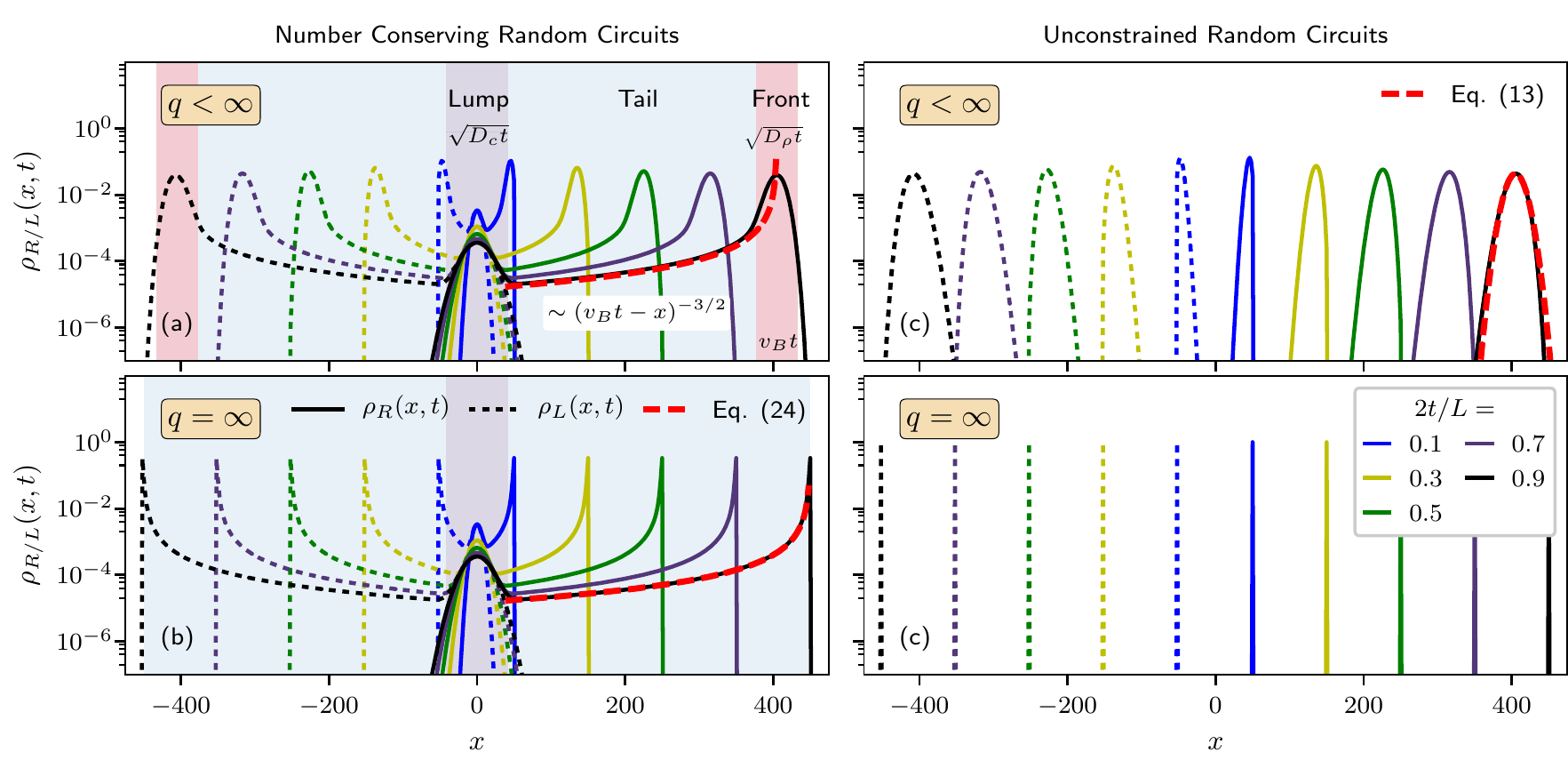}
  \caption{\label{fig:opshape}  (a,b): Right/left-weight profiles $\rho_{R/L}(x,t)$ showing the spreading of an initially local conserved charge $(z\I)_0(t)$ in a random circuit model with $\Sz$ conservation in a system of size $L=1000$ at different times $t$. These profiles depict three regimes: (i)  a ``lump" in the region $|x|\lesssim \sqrt{D_c t}$ reflecting the weight of the  operator on diffusively spreading conserved charges (shaded purple). This lump emits ballistically spreading nonconserved operators at a slow power-law rate.  This emission creates (ii)  the leading ballistic ``fronts" near $|x| \sim v_B t$ 
within which the majority of the operator right- and left-weight is contained (shaded red for the latest time).  These leading fronts are from nonconserved operators emitted at early times and they are perfectly sharp at $q=\infty$ where $\vb=1$ (b), and have a width $\sqrt{D_\rho t}$ for finite $q$ (a); Finally, the slow emission also leads to  (iii)  diffusive tails $\sim(\vb t -|x|)^{-3/2}$ behind the leading fronts which reflect the operator weight in ``lagging'' fronts of nonconserved operator strings that were emitted at later times (shaded blue for the latest time).  The curves in (a) are obtained via a simulation at $q=3$ which takes into account the different processes (diffusion of charges, emission of nonconserved operators and the biased diffusion of the nonconserved right- and left-weights) to order $1/q^2$.  The red dashed curve is the exact infinite $q$ answer for the ``tail'' \eqref{eq:tail}.  (c,d): For comparision, $\rho_{R/L}(x,t)$ in an unconstrained random circuit model\cite{opspreadAdam,opspreadCurt} where $z_0(t)$ isn't ``special". Regimes (i) and (iii) do not exist in an unconstrained circuit, and the ballistically spreading operator fronts describe the entire right- and left-weight profiles. The fronts are again infinitely sharp at $q=\infty$ (d) and have a finite width $\sim \sqrt{D_\rho t}$ for $q< \infty$ (c).}
\end{figure*}

Let us see how this comes about. First, we show in Appendix~\ref{sec:consamps} that the circuit-to-circuit variance in the conserved amplitudes is suppressed both in the  large $q$ and the late time $t$ limit:
\begin{equation}
\Delta_i^{a}(t) \equiv \overline{|a_i^c(t)|^2}  - |\overline{a_i^c(t)}|^2  \sim \frac{1}{q^4}\frac{1}{t^2},
\label{eq:anoise}
\end{equation}
while the leading term $|\overline{a_i^c(t)}|^2$ scales as  $\sim 1/t$  \eqref{eq:adiff}.  In this limit, $\Delta_i^a(t) \approx 0$ and  $\overline{|a_i^c(t)|^2} \approx |\overline{a_i^c(t)}|^2$. Then, the power law decrease in $\rho^c_{\rm tot}$ is obtained as
\begin{equation}
\overline{\rho^c_{\rm tot}(t)} \approx \int dx {|\overline{a^c_x(t)}|^2} = \int dx  \;\frac{1}{2\pi t} e^{-\frac{x^2}{t}} = \frac{1}{2\sqrt{\pi t}},
\label{eq:rho_powerlaw}
\end{equation}
where we coarse-grain at long times to obtain the penultimate equality, using \eqref{eq:adiff}. Likewise, in higher dimensions, this power law decrease will scale as $\rho^c_{\rm tot} (t) \sim t^{-d/2}$ as can be easily seen by considering the higher dimensional generalization of \eqref{eq:adiff}. 
Because of the conservation of total density \eqref{eq:rhoparsedcons}, this decrease in $\rho^c_{\rm tot}$ has to be compensated by a corresponding increase in $\rho^{\rm nc}_{\rm tot}$.  Indeed, from the previous discussion, we see that the local decrease in $\rho^c_{\rm tot}$ (at $q=\infty$) due to action of a gate  $U_{i,\ip}$ \eqref{eq:acircuit}
is 
\begin{equation}
-\delta \rho^c_i(t) =
\frac{(a_{i}(t) - a_{\ip}(t))^2}{2}.
\label{eq:rholocalsource}
\end{equation}  
This is the increase in  $\rho^{\rm nc}_{\rm tot}$ that is locally generated at this gate, and this quantity is proportional to the square of the conserved quantity's local current. These newly-produced non-conserved operators then evolve under the unitary dynamics and get converted to other non-conserved operators with increasing size.  As in the case of the evolution of $\rho_R(x,t)$ for an unconstrained random circuit, the front of non-conserved operators (once generated) spreads ballistically, with $v_B = 1$ for $q=\infty$ since the likelihood of the front moving backwards is again suppressed by $\sim 1/q^2$ (Appendix~\ref{sec:vb}). Further, in the $q=\infty$ limit, there is no ``backflow'' of density from non-conserved operators to conserved charges; this backflow only appears at order $1/q^4$.  Thus, at $q=\infty$ and in the scaling limit, these considerations imply that:
\begin{align}
{\rho_R^{\rm nc}(x,t)} &= -\int_{x-\vb t}^{x} \frac{dy}{v_B} \left. \frac{\partial \rho^c(y,t')}{\partial t'}\right |_{\left(t'=t-\frac{(x-y)}{\vb}\right)} ~.
\end{align}
This expression tells a very natural story. 
The total non-conserved right-weight at position $x$ at time $t$ is the integrated weight of all ``fronts'' emitted at locations $y$ at times $t_y= t- (x-y)/v_B$ such that the front travelling with velocity $\vb$ makes it to position $x$ at time $t$. Moreover, since the conserved charges are primarily spread within a distance $\sim\sqrt{D_c t}$ near the origin, the emission of non-conserved flux is only significant at locations within this diffusively spreading ``lump of charge''. Then, at late time $t$, we see the three pieces in the shape of $\rho_R(x,t)$ mentioned earlier: 
\begin{enumerate}

\item \emph{The diffusive ``lump'':} In the spatial region $|x| \lesssim \sqrt{D_c t}$ near the origin, the right-weight comes almost entirely from the diffusively spreading conserved part of the operator, 
$\rho_R(x,t) \simeq |a^c(x,t)|^2$. 
As a result, the spreading operator has significant weight that is ``left behind'' near the starting position of the operator, and this weight decreases only as a power-law in time $\sim t^{-d/2}$ in $d$ dimensions \eqref{eq:rho_powerlaw}.  By contrast, a spreading operator in the unconstrained circuit model has negligible right-weight (exponentially decreasing with $t$) near its initial location at late times. 

\item \emph{The ballistic front:} The leading ballistic operator front (at the right end) is at $x= \vb t$, and the weight at the leading front is from non-conserved operators that were emitted at early times.  At $q=\infty$, the leading front is sharp, the right-weight is strictly zero for $x> v_B t$ with $v_B=1$, and the sharp front is due to those non-conserved operators that were emitted by gates acting at the precise edge of the causal light cone. 
At finite $q$, as in the unconstrained circuit model\cite{opspreadAdam,opspreadCurt}, the front distributions execute biased diffusion instead of strictly moving forwards at each time step.  This leads to an order $1/q^2$ correction in $v_B$: $\vb \simeq 1- 8/(9q^2)$ (Appendix~\ref{sec:vb}) and gives the main operator front a nonzero width  $\sim\sqrt{D_{\rho}t}\sim \sqrt{t/q^2}$. Thus, at finite $q$, the leading front is mostly due to non-conserved operators that were emitted at early times $t_e \lesssim \sqrt{D_{\rho}t}/v_B$.  For systems in $d>1$, the broadening of the front at finite $q$ is given by a random growth model and grows (if at all) with a smaller power of time\cite{AdamCircuit1,opspreadAdam}. 

\item \emph{The diffusive tail:} Finally, the fronts of non-conserved operator strings that were emitted at later times $t_e \gg \sqrt{D_{\rho}t}/v_B$ lag behind the leading front by $v_B t_e$, leading to the development of power-law tails in the right-weight behind the main operator front.  Consider positions $x$ well separated from both the leading ballistic front and the diffusively spreading charge ``lump'' 
such that $\sqrt{D_c t}\ll x$ and $v_B t-x\gg {\rm max}\{D_{\rho} t, 1\}$.  
In this regime,  at infinite $q$,
\begin{equation}
{\rho_R^{\rm nc}(x,t)} \approx \frac{1}{4 \sqrt{\pi}(t-x/\vb)^{3/2}},
\label{eq:tail}
\end{equation}
which has a power-law tail in both space and time. Indeed, for $x$ far separated from the lump, the diffusive charge lump emitting non-conserved flux can be approximately treated as a point source with the same integrated weight, so that  ${\rho_R^{\rm nc}(x,t)}$ is given by the rate of change of the total conserved density in the lump $\partial \rho^c_{\rm tot}/\partial t \sim t^{-3/2}$ at time $t_e = (t-x/\vb)$, which explains the $3/2$ power \eqref{eq:rho_powerlaw}. The leading functional dependence of this tail is the same at both infinite and finite $q$. For $d>1$ the exponent in this power law is $1+(d/2)$.  By contrast, there is no such tail in the spatial profile of operators evolving under unconstrained circuits, since such circuits have no mechanism for generating fronts that significantly lag behind the main operator front. At $q=\infty$ for the unconstrained circuit,  $\rho_R^{\rm rand} = \delta(x-\vb t)$ so there is strictly no tail; at finite $q$, the right-weight behind the leading front for the unconstrained circuit falls off exponentially in time at locations $x$ behind the front scaling with any fixed $0 \leq x/t < v_B$. 

\end{enumerate} 
The left-weight shows the same three regimes, by reflection symmetry.  

Fig.~\ref{fig:opshape} shows $\rho_{R/L}(x,t)$ for both the unconstrained and charge-conserving random circuit models at infinite and finite $q$ starting with an initially local conserved charge at the origin $(z\I)_0$. The finite $q$ plots are obtained via a simulation at $q=3$ which takes into account the different processes (diffusion of charges, emission of nonconserved operators and the biased diffusion of the nonconserved right- and left-weights) to order $1/q^2$. Higher-order corrections, including back-flow from non-conserved densities to conserved charges, are not included. More explicitly, we use the exact $q$-independent expression for the amplitude of diffusing conserved charges \eqref{eq:abinomial} to obtain both $\rho^c$ and the density of non-conserved operators generated on each bond at every time step \eqref{eq:rholocalsource}. These newly generated non-conserved operators are then assumed to execute biased diffusion, just as in an unconstrained circuit, with a probability of moving backwards equal to $4/(9q^2)$. The net density profile of non-conserved operators is obtained via a spatial and temporal convolution of the ``source'' terms on each bond at each time with the corresponding biased diffusion result for the evolution of $\rho_{R/L}$ in an unconstrained circuit \eqref{eq:rwRand} [with suitably modified expressions for $\vb$ and $D_\rho$, correct to $O(1/q^2)$]. 

\subsubsection{Hydrodynamic description}
We now turn to a long-time hydrodynamic description of the coupled processes involving diffusion of the conserved ``charge'' and the propagation of the fronts of the nonconserved operators emitted from this diffusing conserved charge.  

For specificity, we restrict our attention to systems that are statistically translationally invariant and inversion symmetric.  In such systems the amplitudes of the conserved part of the operator obey an unbiased diffusion equation.  In more than one dimension, the diffusivity may not be isotropic; if that is the case, we rescale distances along the eigendirections of the diffusivity to make it isotropic:
\begin{equation}
\partial_t  {a^c({\bf x},t)}=D_c\nabla^2 {a^c({\bf x},t)}~,
\label{eq:ahydro}
\end{equation}
where $D_c = 1/2$ in our $d=1$ random circuit model \eqref{eq:adiff}. This diffusion is also subject to noise which will lead to fluctuations in $a^c$ across different realizations of the random circuit but, as we discussed above \eqref{eq:anoise}, for initial operators that do contain the conserved quantity the noise 
is a parametrically subleading correction to the amplitudes $a^c$ in the late time limit that is relevant to the hydrodynamics. 

Diffusion is a dissipative process, and this is reflected here in the decrease with time of the total operator weight of the conserved part of the operator, $\rho^c_{\rm tot}(t)=\int d{\bf x}|a^c({\bf x},t)|^2$.  Since the full system is undergoing unitary dynamics, this loss of conserved operator weight means that weight is being converted to nonconserved operator weight.  The density of local rate of this emission of nonconserved operators is $2D_c|\nabla a^c|^2$ \eqref{eq:rholocalsource}, where we show that this coefficient is $2D_c$ below.  Note that this dissipation is a slow hydrodynamic process --- proportional to the conserved quantity's local current squared, as in Ohm's law.  These ``emitted'' nonconserved operators then spread rapidly, becoming highly nonlocal.  This is an example of a presumably much more general picture of how dissipative processes happen within closed systems undergoing unitary dynamics: Correlations captured by low-order observables are moved by the unitary dynamics to highly nonlocal, and thus effectively non-observable operators.  Dissipation in such closed systems is thus the ``hiding'' of correlations in highly nonlocal operators so that the correlations that remain detectable to low-order observables are reduced and thus the ``observable'' entropy increases, even though the von Neumann entropy of the full system remains unchanged.

To more explicitly define the observable entropy, consider the density matrix $\varrho(t) =\frac{1}{(2q)^L} [1+O(t)]$. By unitarity, the von-Neumann entropy of this state is independent of time. Now, if we consider the conserved part of this state $\varrho^c(t) =\frac{1}{(2q)^L} [1+O^c(t)]$, then the entropy of this state is $S^c (t) = \mbox{Tr} [\varrho^c(t) \log \varrho^c(t)]= L\log(2q) - \frac{1}{(2q)^L} \mbox{Tr} [O^c(t)^2 /2]+\cdots$, considering only the leading term in an expansion in powers of $O^c(t)$, which is valid due to the smallness of $O^c$ at late times. Thus, $S^c (t) = S^{\rm eq}_\infty - \rho^c_{\rm tot} (t)/2 +\cdots$. So, $S^c (t)$ increases (towards the infinite temperature equilibrium value) at a rate proportional to the power-law slow rate of decrease of conserved weight  $\rho^c_{\rm tot} (t)$ \eqref{eq:rho_powerlaw}. Since the conserved operators are local (single-site), $\varrho^c(t)$ is ``observable" and $S^c(t)$ is a good proxy for the ``observable" entropy. More generally, one should define the observable part of a state as comprising all local (say single-site) operators rather than just the conserved ones. However, the rate of increase of $S^{\rm obs}(t)$ will still be dominated by the conserved parts since the weight on these only decreases as a power-law in time, in contrast to the exponential in time decrease in the weight of local non-conserved operators. This is the reason why the increase in observable entropy in an unconstrained circuit, while present, is not a hydrodynamically slow process while models with conservation law show a power-law slow increase in the observable entropy at a rate proportional to the local diffusion current \eqref{eq:rholocalsource}.

Next, the dynamics of the spreading of the front of the emitted nonconserved operators is given by a random and nonlinear growth model for $d>1$, which we will not discuss further here\cite{AdamCircuit1,opspreadAdam}. 
For $d=1$ the weighted distributions of the fronts of the spreading strings move by biased diffusion\cite{opspreadAdam, opspreadCurt}.  We will focus on the right-moving front, but the left-moving front is doing the same thing, just spatially reflected so it moves in the opposite direction.

Thus the leading order continuum hydrodynamics in $d=1$ of the right-density $\rho^{\rm nc}_R (x,t)$ is described by a diffusion equation with drift and a source term representing the emission of non-conserved operators from local gradients in $a^c(x,t)$:
\begin{align}
\partial_t {\rho_R^{\rm nc}({ x},t)}&=\vb \partial_x {\rho_R^{\rm nc}(x,t)}+ D_\rho\partial_x^2 {\rho_R^{\rm nc}({ x},t)} \nonumber \\
&+ 2D_c{|\partial_x{a^c({ x},t)}|^2},
\label{eq:rhonchydro}
\end{align}
where we show in Appendix~\ref{sec:vb} that the drift and diffusion constants are $\vb \simeq 1-\frac{8}{9q^2}$ and $D_\rho \simeq \frac{8}{9q^2}$ neglecting corrections of order $1/q^4$ and higher. Further, it was shown in Refs.\cite{opspreadAdam, opspreadCurt} that the circuit-to-circuit fluctuations in $\rho_R(x,t)$ (and hence $\rho_R^{\rm nc}(x,t)$) scale with a parametrically smaller power of $t$, so we can ignore the noise in $\rho_R^{\rm nc}$ in the leading order hydrodynamics. 
Thus, the coupled diffusion equations \eqref{eq:ahydro} and \eqref{eq:rhonchydro} describe the leading order hydrodynamics for those aspects of the shape of the operator that are described by $\rho_R(x,t)$. 

It is instructive to also directly consider the hydrodynamics for the total right-weight, $\rho_R(x,t) = \rho_R^{\rm nc}(x,t) + (a^c(x,t))^2$, an exercise that explicitly reveals 
the form of the source term for $\rho_R^{\rm nc}$ in Eq.~\eqref{eq:rhonchydro}.  Consider
\begin{align}
\partial_t {\rho_R({ x},t)} &= \partial_t {\rho_R^{\rm nc}({ x},t)}  + \partial_t {|{a^c({ x},t)}|^2} \nonumber \\
&= \vb \partial_x {\rho_R^{\rm nc}(x,t)}+ D_\rho\partial_x^2 {\rho_R^{nc}({ x},t)} + \mathscr{S}^{\rm nc}(x,t) \nonumber \\
&+ D_c  \partial_x^2 {|a^c({ x},t)|^2}  - 2 D_c {|\partial_x{a^c({ x},t)}|^2}~, 
\label{eq:rhohydro}
\end{align}
where we've left the functional form of the source term $ \mathscr{S}^{\rm nc}$ for $\rho_R^{\rm nc}$ undetermined, and we've used \eqref{eq:ahydro} in evaluating the time-derivative for the squared amplitudes of the conserved charges. Note however that since $\int dx \rho_R(x,t)$ is conserved, the continuity equation requires that $\partial_t {\rho_R({ x},t)}$ is the total spatial gradient of a current $\mc{J}(x,t)$. Thus, the two terms in \eqref{eq:rhohydro} that are not spatial derivatives must cancel each other (up to a total derivative), giving    
\begin{align}
\mathscr{S}^{\rm nc}(x,t) = 2 D_c {|\partial_x {a^c({ x},t)}|^2}, 
\end{align}
which gives the hydrodynamics for $\rho^{\rm nc}_R$ \eqref{eq:rhonchydro}. 
With this relation, 
\begin{align}
\partial_t {\rho_R({ x},t)} &= \vb \partial_x {\rho_R^{\rm nc}(x,t)}+ D_\rho\partial_x^2 {\rho_R^{nc}({ x},t)} \nonumber \\
&+ D_c  \partial_x^2 {|{a^c({ x},t)}|^2}~, 
\end{align}
which nicely shows that the non-conserved part of $\rho_R$ exhibits diffusive dynamics with a drift, while the conserved part simply diffuses with no net drift. 

\subsection{Spreading of local non-conserved operators}
\label{sec:shape_noncons}
We now briefly describe the spreading of initial operators $O_0$ that are orthogonal to $S_z^{\rm tot}$ such that $\Tr( O_0 S_z^{\rm tot})=0 $. These operators come in two categories: those with $\Delta S_z^{\rm tot} \neq 0$ (such as the raising and lowering operators $r_0$ and $l_0$ with $\Delta S_z^{\rm tot} = \pm 1$ respectively) and those with $\Delta S_z^{\rm tot} = 0$ (such as $r_1l_2$). The first category of operators only have weight on basis strings with the same $\Delta S^z \neq0$ for all times, and thus remain orthogonal to each conserved charge $(z\I)_i$ so that ${a_i(t)}=0 \; \forall \;i, t$. Thus, in this case, $\rho_R(x,t) = \rho_R^{\rm nc} (x,t)$ and these operators do not have any left- or right-weight that is ``left behind'' near the initial location due to slow diffusive dynamics. This means that their right-weight profile does not have diffusive tails and looks essentially the same as $\rho_R$ for an operator  spreading under the action of an unconstrained random circuit \eqref{eq:rwRand}, albeit with different drift and diffusion coefficients. This can also be seen from the hydrodynamic equation \eqref{eq:rhonchydro} since the source term will be zero for this case. However, we will show in the next section that the local operator content \emph{within} the spreading operator, \emph{i.e.} internal to the light-cone, still shows power-law correlations due to the conservation law. 

Let us now turn to the second category of initial operators with $\Delta S_z^{\rm tot} = 0$. These include ``dipole'' operators of the form $O_{01}^{\rm dip} = \frac{(z\I)_0 -(z\I)_1}{\sqrt{2}}$ which still contain the local conserved charges but with amplitudes that sum to zero. In this case, one can show that the coarse-grained Haar-averaged conserved amplitudes $a^c(x,t)$ look like the spatial derivative of the amplitudes obtained for the case starting with a monopole source \eqref{eq:adiff}:
\begin{equation}
\overline{a^c_{\rm dip}(x,t)} \sim  \frac{x}{t^{3/2}} e^{-\frac{x^2}{2t}}, \;\;\; \int dx \; \overline{a^c_{\rm dip}(x,t)} = 0.
\label{eq:adip}
\end{equation}
Indeed, this is also easily obtained from the hydrodynamic diffusion equation \eqref{eq:ahydro} with a dipole initial condition. Then, the total conserved \emph{weight} again decreases as a power law in time, but with a faster decay as compared to the monopole case: $\rho_{\rm tot}^c(t) \sim t^{-3/2}$ in 1d.
The conserved densities again act a source of non-conserved operators leading to a power law tail in $\rho_R(x,t)$, but with scaling $1/(x-\vb t)^{5/2}$ in 1d as is seen by considering the rate of change of $\rho_{\rm tot}^c(t)$. For $q<\infty$, charge-neutral operators like $r_1 l_2$ create a dipole with nonzero probability at early times which then spreads as just described, again giving such a power-law tail in the asymptotic operator shape. 

This brings us to a technical aside about the large $q$ limit which is relevant for the spreading of non-conserved operators.  An initial operator $(r\I)_1 (l\I)_2$ becomes a superposition of order $q^4$ operators, each with $\Delta S^z=0$, under the action of the first gate. Only operators that act as the identity on the qudit spins (and as $z$ on the spin 1/2) have a chance of making a charge dipole and thereby contributing to the conserved weight $\rho^c_{\rm tot}$ --- but these are suppressed in probability by $1/q^4$. This illustrates one aspect of the dynamics that is suppressed by the infinite $q$ limit, namely the likelihood for certain operators to make dipoles and thus to pick up power law tails in $\rho_R(x,t)$.  As a related point, if one starts directly with the dipole operator $O_{01}^{\rm dip}$ as before, then the $q=\infty$ limit is sensitive to whether the dipole is acted upon by a single gate $U_{01}$ at the first time step, or whether it is acted upon by the two gates $U_{-10}$ and $U_{12}$. The latter case proceeds exactly as described above since each initial gate sees a conserved $(z\I)$, while in the former case the single gate immediately converts the dipole to order $q^4$ non-conserved operators leading to a $1/q^4$ suppression in the power-law tail. This strong sensitivity to microscopic initial details in the spreading of this class of operators is a peculiarity of the large $q$ limit. 

\subsection{Spreading of multi-local conserved operators}

We now briefly address the spreading of initial Pauli strings that are a product of $N$ conserved charges, $$O_{i_1 i_2 \cdots i_N} (t=0) = (z\I)_{i_1} (z\I)_{i_2} \cdots (z\I)_{i_N},$$ with the sites ordered so that $i_1<i_2<\dots <i_N$.  The operator dynamics of such strings is also directly constrained by the conversation law since the conservation of $S_z^{\rm tot}$ implies the conservation of $(S_z^{\rm tot})^N = \sum_{j_1 j_2 \cdots j_N}  (z\I)_{j_1}\dots (z\I)_{j_N}$, and thus ``multi-localized'' strings of charges act as the ``conserved densities'' of this higher order conservation law.  Then, as in \eqref{eq:acons}, the sum of all \emph{amplitudes} $a^c_{j_1 j_2 \cdots j_N}$ of the strings of $z$  that appear in the expansion of $(S_z^{\rm tot})^N $ is conserved in time.  Further, if $O(t=0)$ has $N$ conserved charges, then only amplitudes $\overline{a_{\mc{S}}}$ on basis strings with \emph{exactly} $N$ conserved charges survive Haar averaging, even though the expansion $(S_z^{\rm tot})^N$ involves basis strings with less than $N$ charges (in fact the relevant conserved operator is the appropriate weighted sum of $(S_z^{\rm tot})^N$, $(S_z^{\rm tot})^{N-2}$, etc., such that it only contains these basis strings of exactly $N$ charges). 
 
We can now ask about the time evolution of amplitudes of the form $a_{j_1 j_2 \cdots j_N}$ with $j_1<j_2<\dots <j_N$. Note that if all the $j's $ lie on different gates at a given time $t$, then we have independent diffusion of charges on each of the gates and action of the circuit simply averages all the inter-gate correlations and makes them equal. 
Indeed, if one starts in an infinitely large system with the charges in the initial string well-spaced such that the ``diffusive cones'' of the individual charges never intersect, then  $a_{j_1 j_2 \cdots j_N}(t) \simeq \left(\frac{1}{\sqrt{4 \pi D_c t}}\right)^N \prod_{k =1}^N e^{-(j_k - i_k )^2/{4D_c t}}$, and these amplitudes decay with time parametrically faster than those for a single conserved charge. On the other hand, if we start with a finite density of charges such that the independently diffusing charges encounter each other on a gate, then we have to account for the ``hard-core'' interaction between these charges which is encoded in the invariance of $(z\I)_{i}(z\I)_{i+1}$ under the action of a gate. As a result of this, the diffusing charges can never pass each other and the coarse-grained problem is that of ``single-file diffusion'' of $N$ particles with a hard-core contact interaction.  This problem has been solved in the literature\cite{singlefilediffusion}, and predicts that
 \begin{equation}
\overline{ a^c_{j_1 j_2 \cdots j_N}(t)} = \sum_{\sigma \in S_N} \prod_{k=1}^N \frac{1}{\sqrt{4 \pi D_c t}} e^{-(\sigma(j_k) - i_k )^2/{4D_c t}}~,
 \end{equation}
where the sum is over all $N!$ permutations $\sigma$ for the particle labels at time $t$, the initial locations of the charges are $i_1\cdots i_N$, and we require $j_1<j_2 \dots <j_N$. 

Finally, note that initial strings that are products of conserved operators on some sites and non-conserved operators on other sites do not have any overlap with any of the conserved charges or their higher moments.  All Haar-averaged amplitudes for all basis strings will be zero in this case (although two-point correlations of the amplitudes can still survive averaging). Further, even if there is initially a diffusive ``cone'' of locally conserved charges in the vicinity of a conserved operator in the initial string, these quickly lose their coherence upon encountering the ballistic cones emanating from the initial locations of the non-conserved operators. 

\section{Internal structure of spreading operators}
\label{sec:raisingandspin}
So far we have discussed the shape of the spreading operator in terms of the right-weight $\rho_R(x,t)$; we've used this to describe the ballistic ``light cone'' within which the operator has spread, and the power law tails in the distributions of right- and left-weights within the light cone.  We now turn to another layer of structure in the ``shape'' of the spreading operator that is relevant, for example, for the resulting OTOC's.  This is the weighted local distributions of the different qubit operators ($\I, r, l, z$) within the strings $\mathcal{S}$ that make up the operator.  For an unconstrained random circuit, the distribution of these local operators is uniform between the four operator types after the leading operator front has passed a given location, but the distribution is highly biased towards identities before the front reaches.  On the other hand, different operators are inequivalent in the conserving random circuit, leading to imbalances in these local densities that persist after the front passes.  Indeed, the evolution of these local densities and their correlations are constrained by two further conservation laws that encode the fact that the unitary circuit conserves the action of the initial operator under the $U(1)$ symmetry.  These are the conservation of the total ``spin'' $s$ if one starts in a state with definite $S_z^{\rm tot}$, and the conservation of the net ``raising charge'' $\mc{R} \equiv \Delta S_z^{\rm tot}$ of an operator.  Further, we will see that these conservation laws are coupled to each other through suitably-defined two-point correlations of charge/spin within the operator. 

\subsection{Conservation of raising charge}

The action of the conserving circuit preserves the net ``raising charge'' $\mc{R} \equiv \Delta S_z^{\rm tot}$ of an initial operator that starts with a definite $\Delta \Sz$ action. Concretely, for a given basis string $\mc{S}$, the local  $\mc{R}_i = +1$ if the qubit operator in $\mc{S}$ at site $i$ is a raising operator $r$, while a lowering operator $l$ has $\mc{R}_i = -1$, and $z$ and $\I$ have $\mc{R}_i = 0$. Note that $\mc{R}_i$ is neither sensitive to the qudit operator content, nor the operator content on all other sites.  The unitary circuit conserves  $\mc{R}_{\rm tot} = \sum_i \mc{R}_i$ for operators that start with a definite $\mc{R}_{\rm tot}$, and the operator's local ``raising charge'' $\mc{R}_i$ moves diffusively.  Nevertheless, the circuit still converts a given basis string into a superposition of many strings each with different local patterns of $\mc{R}_i$ (but with the same $\mc{R}_{\rm tot}$) and this choice of strings means that the diffusion of ``raising charge'' is subject to noise.  For example, if a unitary gate $U_{12}$ acts on a two-site operator with $\mc{R}_{\rm tot}=+1$ such as $(r\I)_1(z\I)_2$, the action of the gate produces a superposition of $(4q^4)$ two-site strings of the form $\{(ra)_1 (\I b)_2, (ra)_1 (zb)_2, (\I a)_1 (rb)_2, (za)_1 (rb)_2\}$, where $a$ and and $b$ are arbitrary operators on the qudit and, within this ensemble of strings, the raising charge $r$ is equally likely to be on either of the two sites acted upon by the gate (Appendix~\ref{sec:circuitaction}). Of course, there is also noise from circuit-to-circuit fluctuations. If we average over the weighted ensemble of all strings within an operator that is evolved by a particular circuit and across circuits, then 
\begin{align}
\langle \mc R_i(t)\rangle &=\; \overline{\sum_{\substack{\mathclap{\text{strings $\mc{S}:$ } }  \\ \mathclap{\text{$\mc{S}_i = (ra)$}}  }} |a_\mc{S}(t)|^2}\; -\; \overline{ \sum_{\substack{\mathclap{\text{strings $\mc{S}:$ } }  \\ \mathclap{\text{$\mc{S}_i = (la)$}}  }} |a_\mc{S}(t)|^2} \nonumber \\
 &\equiv \overline{\rho_r(i,t)}\; -\; \overline{\rho_l(i,t)}
\label{eq:Rdef}
\end{align}
where $\langle \rangle$ denotes the joint average, and the notation $\mc{S}_i = (ra)$ means that the basis string acts as $r$ on the spin-$1/2$ on site $i$ and acts arbitrarily on the qudit. The last equality defines $\rho_{r/l}(i,t)$ as the weight on all basis strings in the operator expansion of $O(t)$ that locally act as $r/l$ on the spin-$1/2$ on site $i$ (one can analogously define $\rho_{\I/z}(i,t)$). As discussed above, the action of a gate $U_{12}$ makes  $\langle \mc{R}_i\rangle $ equal on the two sites and produces a ``raising charge'' current $\sim \partial_x \mc{R}(x,t)$:
\begin{align}
\langle \mc R_1(t+1)\rangle = \langle \mc R_2(t+1)\rangle = \frac{\langle \mc R_1(t)\rangle +\langle \mc R_2(t)\rangle}{2},
\label{eq:Rcircuit}
\end{align} 
which is identical to the structure we had previously obtained for the dynamics of $\overline{a_i^c(t)}$ \eqref{eq:acircuit}. Thus, any initially local spatial distribution of $\langle\mc{R}_i\rangle$, for example if $O_0 = (r\I)_0$, spreads diffusively with diffusion constant $D_r = D_c$. 

\subsection{Conservation of spin}

Likewise, there is a noisy diffusion process governing the conservation of total spin $s$ which measures the projections onto definite $S_z^{\rm tot}$ states. If we rotate the local spin-1/2 operator basis to the projection ``up'', $u = (\I + z)/\sqrt{2}$ and ``down''  $d = (\I - z)/\sqrt{2}$ operators, then these are charged under $s$ as $s=\pm 1$ respectively, while $r,l$ have $s=0$. In this basis, the operators $(u\I)_i(u\I)_{\ip}$, $(d\I)_i(d\I)_{\ip}$  and $\frac{(u\I)_i(d\I)_{\ip}+ (d\I)_i(u\I)_{\ip}}{\sqrt{2}}$ are special and left invariant by the action of all two-site gates $U_{i,\ip}$, while the others mix between themselves in a manner that locally conserves $s_{\rm tot} = \sum_i s_i$ on each gate. Thus, as before, the circuit conserves total $s_{\rm tot}$ while generating noise, so that after averaging over circuits and strings 
\begin{align}
\langle  s_i(t)\rangle &=\; \overline{\sum_{\substack{\mathclap{\text{strings $\mc{S}:$ } }  \\ \mathclap{\text{$\mc{S}_i = (ua)$}}  }} |a_\mc{S}(t)|^2}\; -\; \overline{ \sum_{\substack{\mathclap{\text{strings $\mc{S}:$ } }  \\ \mathclap{\text{$\mc{S}_i = (da)$}}  }} |a_\mc{S}(t)|^2} \nonumber \\
 &\equiv \overline{\rho_u(i,t)}\; -\; \overline{\rho_d(i,t)}. 
 \label{eq:sdef}
\end{align}
The action of a gate makes the average spin equal on the two sites:
\begin{align}
\langle  s_1(t+1)\rangle = \langle  s_2(t+1)\rangle = \frac{\langle  s_1(t)\rangle +\langle  s_2(t)\rangle}{2}.
\label{eq:scircuit}
\end{align} 
This again means that an initial spin polarization spreads diffusively with $D_s = D_c$.  Note that this equality between diffusivities and the one for raising charge above \eqref{eq:Rcircuit} are not completely trivial statements, since $D_c$ is for the {\it amplitudes} of the conserved operators that are a non-identity only at one site, while $D_s/D_r$ are for the local operator {\it weights} (amplitudes squared) of $u/d$'s and $r/l$'s within {\it all} strings, both conserved and non-conserved.

\subsection{Coupling between spin and raising charge}

If the initially local operator contains a nonzero net value of either of these two ``charges'', then that charge will only spread diffusively.  Thus, if we look near the ballistically-moving operator front at long times, none of this initial charge can be there yet.  Moreover, the action of the circuit  is symmetric with respect to interchanging $r \leftrightarrow l$ and $u \leftrightarrow d$ (Appendix~\ref{sec:circuitaction}), so that no new imbalances of the average charges is created due to the action of the circuit. Thus, both these average charges are zero far from the initial location at late times so that $\rho_r(x,t) = \rho_l(x,t)$ and $\rho_u(x,t) = \rho_d(x,t)$  for $|x| \gg \sqrt{D_c t} $. 
However, for all ballistically spreading operators,  there is a next layer of structure in the\emph{ two-point} correlations of these charge densities that evolves diffusively near the front and influences some of the OTOC's. This structure stems from the initial condition in which the local operators are the identity on all sites except those near the origin, and these identities only contribute to $\rho_{u/d}$ and not to $\rho_{l/r}$. That is,  \emph {before} the front comes through, the operator is locally the identity which is an equal-amplitude linear combination of all strings that contain only $u$'s and $d$'s.  On the other hand, in the final ``local equilibrium'' state of the operator, all local strings are equally likely a long time after the front has passed through (up to corrections that decay as a power law in $L$). This equilibrium state is reached via a noisy diffusion process that turns on at each position when the front passes through that position.  The two point correlations before the front passes through are $\langle \mc R_i \mc R_j \rangle=0$ and $\langle s_i s_j \rangle=\delta_{ij}$, and these serve as the initial conditions on the diffusion process. This diffusion relaxes the initial imbalances between the \emph{relative} density of local operators charged under spin and raising charge, in particular the imbalance in $(\rho_u(x,t) + \rho_d(x,t)) - (\rho_r(x,t) + \rho_l(x,t))$.  We now see how this comes about. 

To start, we look at the dynamics of \emph{inter}-gate correlations.  Consider two different gates acting at the same time $t$ on sites $i$, $i+1$ and $j$, $j+1$.  These two gates simply set all inter-gate correlations of each type equal, just as they do for the average densities:
\begin{align}
\langle  s_i s_j \rangle(t+1) = \frac{\langle  s_i s_j +  s_i s_{j+1} + s_{i+1} s_j +  s_{i+1} s_{j+1}\rangle(t)}{4}~,
\label{eq:scorrinter}
\end{align}
and 7 other similar equations for the other inter-gate correlations of $s$ and $\mc R$ of the form $\langle  \mc R_i \mc R_{j+1} \rangle(t+1)$ etc. For concreteness, $  s_i s_j = +1$ for strings that locally look  like $u_i u_j/ d_i d_j$ on sites $i,j$ and $ s_i s_j = -1$ for strings that look like $u_i d_j/ d_i u_j$, and is zero otherwise (analogous expressions for  $\mc{R}_i \mc{R}_{j}$), and the equation above can be derived by looking at the action of the circuit gates on the different types of operators (Appendix~\ref{sec:circuitaction}).
The zero inter-gate correlations before the front comes through are indeed invariant under this dynamics.  The coarse-grained diffusion equation for this two-point correlation for $x \neq y$ is thus
\begin{align}
\partial_t \langle s(x)s(y)\rangle(t) &=  D_c(\partial_x^2 + \partial_y^2)\langle s(x)s(y)\rangle(t) ~,
\label{eq:diff}
\end{align}
with $\langle \mc R(x)\mc R(y)\rangle(t)$ obeying the same equation.  This is for $d=1$, but the generalization of this and the results below to higher $d$ seems straightforward.

Thus far we have two separate diffusion processes governing the diffusion of ``raising charge" $\mc R$ and ``spin" $s$ correlations.  In fact, these two processes are coupled once we consider the action within one gate on ``dipoles''.
The ``dipoles'' of $\mc{R}$ are net raising-neutral operators like $\{(ra)_i(lb)_{\ip}, (la)_i(rb)_{\ip}\}$, while the dipoles of spin $s$ are operators of the form $\{(ua)_i(db)_{\ip}, (da)_i(ub)_{\ip}\}$.  The coupling between spin and raising charge stems from the fact that while the total density of all such dipoles within one gate is locally conserved, the different dipole species mix among each other under the action of the gate.  And it is this process that allows an initial operator that contains no $r$'s or $l$'s to relax to the final equilibrium where all local strings are equally likely.  The precise ``boundary conditions'' that couple the above diffusion equations \emph{within} one gate \emph{i.e.} at $x=y$ are $q$-dependent and involve other correlations that can initially be well out of equilibrium just after the front passes. 

\begin{figure}
  \includegraphics[width=\columnwidth]{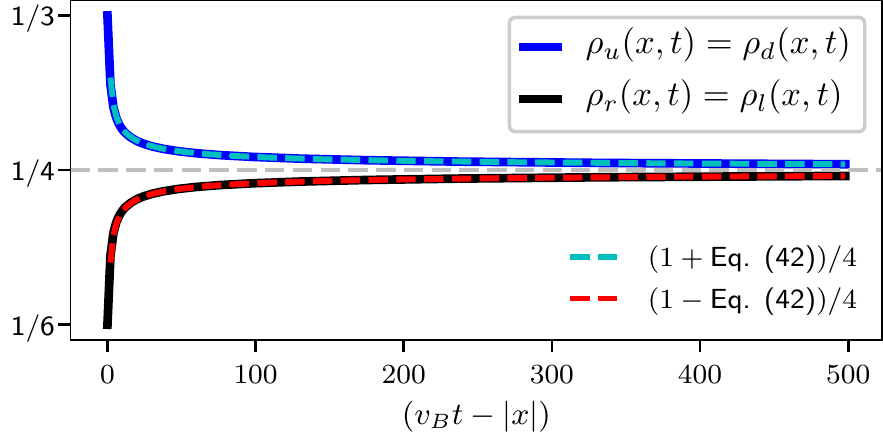}
  \caption{\label{fig:Cij}  Densities of local operators charged under ``spin'' and  ``raising action'' behind the ballistic front of a spreading operator at $q=\infty$ and late times \eqref{eq:deltars_exact}.  $\rho_{u/d}(x,t)$ measures the local operator weight on site $x$ on $u/d$ operators that are charged under spin, while $\rho_{r/l}(x,t)$ measures the local weight on $r/l$ operators that have raising charge. Outside the ballistic operator front (at $|x|>\vb t$), the spreading operator locally is purely identities which contribute equally to $\rho_{u/d}=1/2$, but do not contribute to $\rho_{r/l}=0$.  The arrival of the front at a given site turns on a noisy coupled diffusion process between the spin and raising charges which relaxes the initially imbalanced densities of these charges to the final ``equilibrium'' value where all local operators are equally likely. Note that $\rho_u = \rho_d$ and $\rho_r = \rho_l$ in this regime since any initial imbalances in raising/spin charges spread only diffusively and are not present near the ballistic front at late times. Dashed lines plot the  ``coarse-grained" densities in the scaling limit \eqref{eq:deltars}.
  }
\end{figure}

The coupling between the charges can be simplified by looking at particular linear combinations of the $\langle \mc{R}_i\mc{R}_j\rangle $ and $\langle s_i s_j\rangle$ correlations. 
If we consider $\langle \mc R_i \mc R_j + s_i s_j \rangle$, this has the initial condition $\langle \mc R_i \mc R_j + s_i s_j \rangle=\delta_{ij}$.  This initial condition is identical to the equilibrium final state where all strings are equally likely.  And if we consider the action both within one gate and between two different gates, it does not change this quantity, which is therefore time-independent. Thus, although we do have two coupled diffusion equations, it reduces to just one diffusion equation for 
\begin{equation}
G_{ij}(t) \equiv \langle  s_i s_j - \mc R_i \mc R_j\rangle(t)
\label{eq:Cij}
\end{equation}
for this initial state, which also has $G_{ij}(0)=\delta_{ij}$ as its initial condition.  The local effective diffusivity due to the processes within one gate is, in general, $q$-dependent and different from that when $i$ and $j$ are in different gates, but at long time this affects a ``region'' of $G_{ij}$ that is a fraction $\sim \sqrt{1/t}$ of the distance $(i-j)$ over which it has spread.  Thus the leading coarse-grained long-time behavior of $G_{ij}$ is
\begin{align}
G(x,y,t) \approx  \frac{1}{2\sqrt{\pi t}} e^{-\frac{(x-y)^2}{4t}}~,
\label{eq:cijt}
\end{align}
where the time $t$ here is measured from the time when the front passes, say, the point $(x+y)/2$.  Note that the effective diffusivity here is $2D_c=1$, since we have independent diffusion at the locations $x$ and $y$. In fact, at $q=\infty$, the evolution of $G_{ij}(t)$ within one gate is identical in structure to the intra-gate evolution and we can solve for $G_{ij}$ exactly in this limit (Appendix~\ref{sec:Cij}), and the coarse-grained answer agrees with \eqref{eq:cijt}.  Finally, note that this form for $G(x,y,t)$ when evaluated at $x=y$ shows how the initial imbalances in the populations of $u,d$ vs. $r,l$ decay diffusively away from the location of the front:
\begin{align}
\Delta_{\mc{R}s}(x,t) &\equiv G(x,x,t) \nonumber \\
&= (\rho_u(x,t) + \rho_d(x,t)) - (\rho_r(x,t) + \rho_l(x,t)).
\end{align}
 For $q=\infty$, where $\vb =1$, these imbalances can be evaluated exactly and they take the form 
\begin{align}
\Delta_{\mc{R}s}(x,t) 
&= \frac{1}{3}\sum_{n= \floor{\frac{( t-|x|)}{2}}}^{2\floor{\frac{( t-|x|)}{2}}}\left(\frac{1}{3}\right)^{n} {n \choose 2n - 2\floor{\frac{( t-|x|)}{2}}}\times  \nonumber \\
 &\qquad \qquad \qquad \qquad \quad \;\;{2n - 2\floor{\frac{( t-|x|)}{2}} \choose  n - \floor{\frac{( t-|x|)}{2}}    }\label{eq:deltars_exact} \\
&\approx  \frac{1}{2\sqrt{\pi (t-|x|/\vb)}}.
\label{eq:deltars}
\end{align}
where we've taken the late-time coarse-grained limit in the last line to evaluate the sum via a saddle-point approximation (Appendix~\ref{sec:Cij}), and restored units.  At finite $q$, we expect the leading functional dependence to still be the same, just with a reduced $\vb$.  This equation, coupled with the fact that $\rho_l \approx \rho_r$ and $\rho_u \approx \rho_d$ away from the diffusive lump allows us to solve for the local densities of the different operators, and this is sketched in Fig.~\ref{fig:Cij}, showing how all local densities diffusively approach the equilibrium value of $1/4$ after the front passes through. This diffusive relaxation of the imbalance leads to a power-law diffusive tail in certain OTOC's, as we will see in the next section. 
   
\section{Out-of-Time-Order Commutators}
\label{sec:otoc}

We now turn to measuring the out-of-time-order commutator between different classes of operators~\eqref{eq:otoc}.  We  find that, in several cases, this quantity is sensitive to both the ``shape'' and the internal structure of spreading operators. If we consider the OTOC between two generic (initially) local operators, then this quantity is close to zero until the arrival of the ballistically spreading operator front and shows a sharp increase upon the arrival of the front. However, the OTOC for systems with local conservation laws also generically develops a ``diffusive tail'' and  approaches its late time asymptotic value only as a power-law in time. Let us now see how this arises. 

Define 
\begin{align}
\mc{C}_{\alpha \beta}^{\mu}(x,t) &= \frac{1}{2} \Tr \;\{\rho_\mu^{\rm eq}  |[\sigma_0^\alpha(t), \sigma_x^\beta]|^2\}~,
\end{align}
where $\sigma_x^{\{\alpha = 1,2,3\}} = \{r_x,l_x,z_x\}$ as before, we suppress the identity operators acting on the qudit spins on sites $0,x$ for notational simplicity, and 
\begin{equation}
\rho_\mu^{\rm eq}=\frac{e^{\mu S_z^{\rm tot}}}{\mbox{Tr} \; e^{\mu S_z^{\rm tot}}}
\end{equation}
is the equilibrium ``grand-canonical'' ensemble at a particular chemical potential $\mu$. Note that while the infinite-temperature average is the only equilibrium for a random circuit model with no energy conservation, for a circuit model with an extra conservation law (like ours), we may also consider a non-zero chemical potential which weights the different spin sectors as above. We will focus on the equal-weight ensemble with $\mu=0$  for the majority of this section, briefly commenting on $\mu \neq 0$ towards the end. 
We now study the OTOC for several different choices of $\alpha, \beta$.

\subsection{OTOC between $z_0(t)$ and $r_x$}
Let us start with $\mu = 0, \alpha = 3=z$ and  $\beta=1=r$. Then 
\begin{align}
\mc{C}_{zr}^{0}(x,t) &= \frac{1}{2} \Tr \; \{\rho_0^{\rm eq}  |[z_0(t), r_x]|^2\}~,
\end{align}
and it is clear that only basis strings in the operator expansion of $z_0(t)$  that act as $(la)$ or $(za)$ on site $x$ contribute to $\mc{C}_{zr}^0(x,t)$. Then, using the commutation relations  $[z_x,r_x]=2r_x$ and  $[l_x,r_x]=-2z_x$, and the orthonormality of basis strings, we get
$$\mc{C}_{zr}^{0}(x,t) = \sum_{\substack{{\text{strings $\mc{S}:$ } }  \\ \mathclap{\text{$\mc{S}_x = (la),(za)$}}  }} 2|a_\mc{S}^z(t)|^2$$
where the notation $a_\mc{S}^z(t)$ denotes the amplitude of basis string $\mc{S}$ in the  expansion of $z_0(t)$. A bit of algebra shows
\begin{align}
 \mc{C}^{0}_{zr}(x,t) &=& 1&\;\;\;\;&-&\;\;\;\; \sum_{i< x} \rho_R(i,t) \nonumber \\ 
&+&\sum_{\substack{\mathclap{\text{strings $\mc{S}:$ } }  \\ \mathclap{\text{ RHS$(\mc{S})\geq x$ and}}\\ \mathclap{\text{ $\mc{S}_x = (\l a),(z a)$}}  }}& |a^z_\mc{S}|^2
\;\;\;\;&-&\;\;\;\;
\sum_{\substack{\mathclap{\text{strings $\mc{S}:$ } }  \\ \mathclap{\text{ RHS$(\mc{S})\geq x$ and}}\\ \mathclap{ \text{ $\mc{S}_x = (\I a),(ra)$}}  }} |a^z_\mc{S}|^2, 
\label{eq:Czr_parsed}
\end{align}
which is a useful way of writing this for $x>0$; an analogous expression involving the left-weight $\rho_L$ can be written for $x<0$. 

\begin{figure}
  \includegraphics[width=\columnwidth]{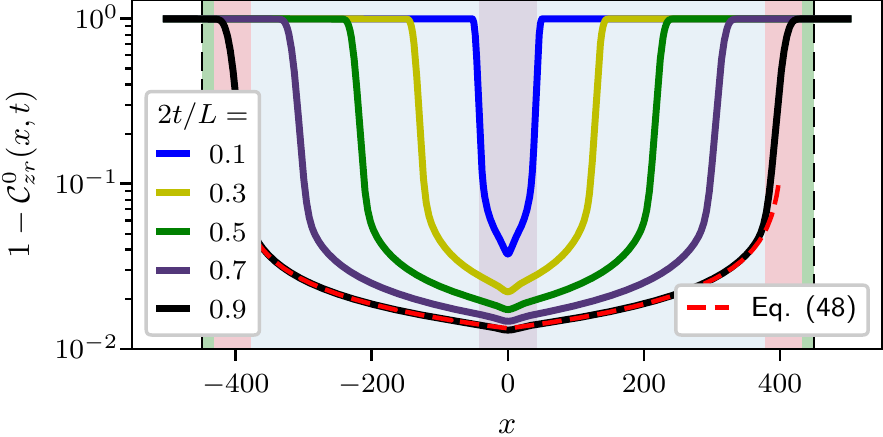}
  \caption{\label{fig:otoc}  One minus the out-of-time-order commutator (OTOC) between $z_0(t)$ and $r_x$ at zero chemical potential, $\mc{C}^0_{zr}$, plotted against $x$ for a system of length $L=1000$ at different times $t$ showing the different regimes discussed in the text. 
  For $|x| >t $ (outside the dashed vertical lines), the OTOC is strictly zero due to the locality of the circuit.  In the region $\vb t <|x|<t$, which is inside the causal light cone but before the leading front arrives, the OTOC is exponentially small (green shaded area for the latest time). The arrival of the ballistic operator front ($|x| \sim \vb t$) leads to a strong increase in the OTOC from a value exponentially small to an $O(1)$ value (shaded red area for the latest time).  However,  diffusive tails in the operator shape or internal structure lead to diffusive power-law tails in space and time $\sim (x-\vb t)^{-1/2}$ in the late-time approach of the OTOC to its final value of 1 (shaded blue area for the latest time).  By contrast, for an unconstrained random circuit (not shown), the OTOC at a given site approaches one exponentially quickly after the leading front passes\cite{opspreadAdam,opspreadCurt}. The diffusive region near the origin $|x| \lesssim \sqrt{D_c t}$ (shaded purple) receives a subleading $1/t$ contribution from the conserved charges which shows up as a ``dimple'' in the curves at early times which becomes weaker at late times. All curves are obtained via a simulation using $q=3$ and taking into account all processes to order $1/q^2$. The dashed red curve is the $q=\infty$ prediction for the functional form of the tail \eqref{eq:otoc_tail}.  }
\end{figure}

Note that the first two terms in \eqref{eq:Czr_parsed} are sensitive to the overall operator shape, namely the right-weight profile of the spreading operator, while the latter two terms are sensitive to the local ``internal'' operator content after the front has passed site $x$. In random circuits with no conservation laws, all operator types are equally likely after the front has passed through and thus only the first two terms contribute substantially to the OTOC\cite{opspreadAdam,opspreadCurt}.  Notice that for this particular pair of operators, the second line measures the local raising charge $\mc{R}_x$ ~\eqref{eq:Rdef} and the difference between the local densities of $z$'s and $\I$'s at site $x$ after the front has passed $x$. 
As discussed in the previous section, there is no average raising charge away from the diffusing ``lump" near the origin, and the difference in local densities of $z$'s vs. $\I$'s is also zero \emph{except} near the origin and right at the front. So, to leading order the second line in \eqref{eq:Czr_parsed} vanishes in the region between the diffusive lump and the front.
Further, since we start with zero raising charge, the only contribution to the second line comes from the imbalance between $\rho_z(x)$ and $\rho_\I(x)$. Indeed, all conserved operators $(z\I)_i$ with $i \geq x$ contribute  to $\rho_\I(x)$, while only $(z\I)_x$ contributes to  $\rho_z(x)$. These conserved operators are the leading source of the imbalance between $\rho_\I(x)$ and $\rho_z(x)$ near the origin. 
Non-conserved strings do not substantially contribute to the imbalance between $\rho_z(x)$ and $\rho_\I(x)$ away from the front.  Then, putting it all together, 
\begin{align}
1 - \mc{C}^{0}_{zr}(x,t) &\approx \sum_{i< x} \rho_R^{\rm nc}(i,t) + \sum_{i> x} \rho_L^{\rm nc}(i,t)   \nonumber \\ 
&\;\;\; + \rho_{\rm tot}^c(t)-2\rho^{\rm c} (x, t)~,
\label{eq:Czr_parsed2}
\end{align}
where we've separated the contributions from conserved and non-conserved operators \eqref{eq:rhoparsed} and used also the left-weight $\rho_L^{\rm nc}$ to produce an expression that is valid at late time for all $x$.
Since $\rho_{\rm tot}^c(t) \sim t^{-1/2} \gg \rho^{\rm c} (x, t)\sim t^{-1}$ at late times, the contribution of $\rho^{\rm c} (x, t)$ is also subdominant at late times and will be neglected below, although it is visible as a weak ``dimple'' in the diffusive regime near the origin in Fig.~\ref{fig:otoc}.  Thus, away from the fronts and the diffusive regime near the origin, the OTOC for this pair of operators receives its primary contribution from the total right- or left-weights alone, just as in the unconstrained circuit\cite{opspreadAdam, opspreadCurt}.  We saw above that the right- and left-weight profiles for $z(t)$ show diffusive power-law tails, and these translate into diffusive tails for this OTOC as well. 

Let us now examine this OTOC in different regimes at late times such that the leading ballistic front is well separated from the diffusive ``lump'' near the origin: $\sqrt{D_c t} \ll \vb t$. 
\begin{enumerate}

\item \emph{Outside the ``light cone'' $|x|>t$}: Due to the locality of the circuit, a spreading operator $O_0(t)$ acts as the identity outside the light cone defined by $|x| \leq t$. Thus, the commutator  $\mc{C}(x,t)=0$ in this regime. 

\item \emph{Beyond the leading operator fronts so $|x| \gg \vb t+ \sqrt{D_{\rho} t}$, but within the causal light cone so $|x|<t$}:  Before the main operator front gets to $x$,  any operator weight that is not locally the identity is exponentially small in $t$ and thus the OTOC is also exponentially small in $t$ (for fixed $x/t$ in this regime).  This regime does not exist for $q=\infty$. Ref.~\onlinecite{opspreadCurt} showed that the OTOC shows a near-exponential increase with time in this regime for the unconstrained circuit model, but with a position-dependent analog of the ``Lyapunov exponent".  We expect the same qualitative behavior in this regime for our model, since this operator ``edge'' just comes from non-conserved operators emitted at early times whose right- and left-weights then show biased diffusion, just as in the unconstrained circuit. 

\item \emph{Within the leading operator fronts, $|x|-\vb t \sim \sqrt{D_{\rho} t}$}:  This regime describes the growth of the OTOC from an exponentially small value in $t$ to an order one number due to the arrival of the ballistic front. Here the operator right- or left-weight is again from non-conserved operators emitted at early times and, to leading order at long time for finite $q$, is a Gaussian of width $\sim \sqrt{D_{\rho} t}$, so the leading behavior of the OTOC in this regime is given by the corresponding error function that is the integral of this Gaussian profile, just as in the unconstrained circuit\cite{opspreadAdam, opspreadCurt}. 

\item \emph{In the ``tails'', $  |x| \ll \vb t -  \sqrt{D_{\rho} t} $}: This regime describes the late-time approach of the OTOC to its final asymptotic value long after the main front has passed site $x$. In this regime, the deviation of the OTOC from its final value of one is given by the total weight (conserved and non-conserved) of operator strings that have not yet reached site $x$ at time $t$.  This is obtained by considering the total conserved weight at time $t' = (t-|x|/\vb)$, since any non-conserved fronts emitted after after time $t'$ do not reach site $x$ by time $t$. Then, from \eqref{eq:Czr_parsed2} and \eqref{eq:rho_powerlaw},
\begin{equation}
1 - {C}_{zr}^{0}(x,t) \approx \frac{1}{2\sqrt{\pi(t-(|x|/\vb))}}.
\label{eq:otoc_tail}
\end{equation}
Thus, the tails in the right- and left-weights of $z_0(t)$ lead to tails in the OTOC and, long after the front has passed site $x$, the OTOC still has a power law (in $x$ and $t$) deficit from its asymptotic value.  By contrast, in an unconstrained random circuit, the OTOC approaches 1 exponentially in time after the front passes.  This slow power-law approach characterizes the ``diffusive tail'' in the late time behavior of the OTOC.  In the diffusive regime near the origin, $|x| \sim \sqrt{D_c t}$, there are further corrections to this of order $\sim 1/t$ from $\rho_c(x)$ that smoothly connect the OTOC's $x$-dependence between $x<0$ and $x>0$.

\end{enumerate}
Fig.~\ref{fig:otoc} shows a sketch of ${C}_{zr}^{0}(x,t)$ at different times, depicting the different regimes above. 

\subsection{OTOC between $r_0(t)$ and $z_x$}
Let us now consider the OTOC between $r_0(t)$ and $z_x$ so that $\mu = 0, \alpha = 1=r$ and  $\beta=3=z$. Then 
\begin{align}
\mc{C}_{rz}^{0}(x,t)  
&=& 1&\;\;\;\;&-&\;\;\;\; \sum_{i< x} \rho_R(i,t) \nonumber \\ 
&+&\sum_{\substack{\mathclap{\text{strings $\mc{S}:$ } }  \\ \mathclap{\text{ RHS$(\mc{S})\geq x$ and}}\\ \mathclap{\text{ $\mc{S}_x = (r a),(l a)$}}  }}& |a^z_\mc{S}|^2
\;\;\;\;&-&\;\;\;\;
\sum_{\substack{\mathclap{\text{strings $\mc{S}:$ } }  \\ \mathclap{\text{ RHS$(\mc{S})\geq x$ and}}\\ \mathclap{ \text{ $\mc{S}_x = (\I a),(za)$}}  }} |a^z_\mc{S}|^2, 
\label{eq:Crz_parsed}
\end{align}
In this case, the first line is again sensitive to the overall operator shape, while the second line cares about the relative density difference within strings between local operators that are charged vs. uncharged under $\mc{R}$, i.e. the difference in the local densities of $r/l$'s compared to $\I/z$'s --- note that this is a different combination of local densities as compared to the one in \eqref{eq:Czr_parsed}. 

Next, note that the averge OTOC between $r_0(t)$ and $z_x$ should equal the average OTOC between $z_0(t)$ and $r_x$ since the two are related by time-reversal:
$\mc{C}_{rz}^{0}(x,t)  = \mc{C}_{zr}^{0}(x,-t)$,  and the average properties of our circuit are invariant under time-reversal. In the previous subsection, we saw that the diffusive tail in the OTOC $\mc{C}^0_{zr}$ \eqref{eq:otoc_tail} came from the tail in the right-weight profile of $z_0(t)$. However, recall from Section~\ref{sec:shape_noncons} that the spreading $r_0(t)$ shows no tails in its $\rho_R$, since this operator starts out orthogonal to the conserved charges so that $\rho_R(x,t) \sim e^{-(x-\vb t)^2/4D_\rho t}/\sqrt{4\pi D_\rho t}$, and $\rho_R(x,t)$ is exponentially small for $x \ll \vb t  $. Thus, the tail in $\mc{C}^0_{rz}$ behind the ballistic front must come from the imbalance between the density of raising charges $r/l$ and spin charges $u/d$ (which are superpositions of $\I/z$) . Indeed, we saw in Section~\ref{sec:raisingandspin} that the coupling of spin and raising charge combined with the initial condition at the left and right fronts leads to a diffusive decay of the imbalance $\Delta_{\mc{R}s}$ between the local $\rho_r/\rho_l$ and $\rho_u/\rho_d$ densities away from the locations of the fronts \eqref{eq:deltars} (Fig.~\ref{fig:Cij}). Using \eqref{eq:deltars}, we find that the OTOC  $\mc{C}^0_{rz}$ in the regime $ \sqrt{D_c t}\ll |x| \ll \vb t  $ is given by $1 - \mc{C}^0_{rz}(x,t) \approx \Delta_{\mc{R}s}(x,t) \approx \frac{1}{2\sqrt{\pi (t-|x|/\vb)}}$, which agrees with \eqref{eq:otoc_tail}, as it must. 

This case provides a nice example of how the OTOC can probe the \emph{internal} operator content inside the light cone and show diffusive tails, even when there are no tails in the overall operator shape of the spreading operator as measured by $\rho_{R/L}$. 
\\
\subsection{OTOC between $z_0(t)$ and $z_x$}
Turning next to the ``diagonal'' case $\mu = 0, \alpha = \beta = 3=z$, we find that $\mc{C}^0_{zz}$ has the same structure as in \eqref{eq:Crz_parsed}, but $\rho_R$ now refers to the right-weight profile of $z_0(t)$.  Thus, this commutator receives contributions \emph{both} from the tails in the right-weight and from the tails in the imbalance between the spin and raising charges, so that $1 - \mc{C}^0_{zz}(x,t) \approx \frac{1}{\sqrt{\pi (t-|x|/\vb)}}$ for $|x| \ll \vb t$.  Thus, a commutator between two conserved charges couples more strongly to the diffusive processes in the system, doubling the amplitude of the diffusive power-law tail in the OTOC, relative to the two other cases discussed above.  

\subsection{OTOC between $r_0(t)$ and $r_x$}
Finally, consider the commutator between $r_0(t)$ and $r_x$ so that $\alpha = \beta = 1$. Note that $\mc{C}^0_{rr}$  has the same structure as in \eqref{eq:Czr_parsed} (with $\rho_R$ referring to the right-weight of $r_0(t)$). This commutator again shows a sharp increase when the ballistic front reaches site $x$, but now \emph{both} the contributions from the right-weight and the ``internal'' operator content are exponentially suppressed away from the front so this OTOC does \emph{not} show power law tails in the regime $\sqrt{D_c t} \ll x \ll \vb t$ and approaches its asymptotic value exponentially after the front passes through. Instead, if we probe the commutator at short distances $x \sim \sqrt{D_c t} $, then the OTOC displays a weak effect within the ``diffusive cone'' from the diffusion of the initial raising charge \emph{i.e.} the diffusion of the imbalance between $\rho_r$ and $\rho_l$. The same is true for OTOC's between $r(t)$ and $l_x$. 

This result emphasizes that the diffusive tails in the OTOC at long distances and late times are tied to the operator dynamics of the conserved charges in the system. OTOC's between operators that are both orthogonal to the conserved charges do not display diffusive tails behind their fronts.  Of course, if one simply measures the OTOC between two generic local operators, then these will have some overlap with the conserved charges and show diffusive tails. 

\subsection{ $\mu \neq 0$}
We now consider OTOC's evaluated in an equilibrium ensemble with a chemical potential $\mu >0$ that weights different $S_z^{\rm tot}$ sectors , and we will be interested in understanding the degree to which this compares with the finite temperature OTOC's that have been evaluated for Hamiltonian models.  A  difference between the energy conserving and  $\Sz$ conserving cases is that, on the Hamiltonian side, both the Gibbs factor $e^{-\beta H}$ and the time-evolution are governed by the same Hamiltonian. On the other hand, for the $U(1)$ problem, the ``Gibbs factor'' $e^{\mu N}$ commutes with the time evolution operator $U(t)$ but is otherwise unrelated to it. This distinction between the two becomes important when considering non-zero $\mu$.  

To start, consider again the equilibrium Gibbs ensemble for a given chemical potential: 
\begin{align}
\rho_\mu^{\rm eq}&=\frac{e^{\mu S_z^{\rm tot}}}{\mbox{Tr} \; e^{\mu S_z^{\rm tot}}} = \prod_{i=1}^L \frac{1+z_i \tanh{\mu}}{2q }\nonumber\\
&= \prod_{i=1}^L \frac{1}{2q}\left[ u_i \left(\frac{1+\tanh{\mu}}{\sqrt{2}}\right)+ d_i \left(\frac{1-\tanh{\mu}}{\sqrt{2}}\right)\right], 
\label{eq:rhomu}
\end{align}
where we've switched to the  (normalized) $u/d$ basis which projects onto up and down conserved spins in the last line. In the limit  $\mu \rightarrow \infty$, 
\begin{equation}
\rho_\infty^{\rm eq} = (2q)^{-L} \prod_i (\sqrt{2}  u_i) = q^{-L} \prod_i (|\uparrow \rangle\langle \uparrow|\otimes  \I )_i.
\end{equation}
In this limit, $\rho_\infty^{\rm eq}$ projects the conserved spins to the ``all up'' state so that the conserved part of the dynamics is completely ``frozen out", reducing the system to just the unconstrained qudit spins. The unconstrained qudit system is chaotic and displays ballistic growth of operators with a non-zero $q$-dependent butterfly velocity, just as in the unconstrained random circuit model\cite{opspreadAdam, opspreadCurt}. This illustrates how the independence of $U(t)$ and $\rho^{\rm eq}$ can lead to chaotic dynamics even in the $\mu \rightarrow \infty$ limit, while the analogous $T \rightarrow 0$ limit for Hamiltonian systems arrests chaos.  On the other hand, if one considers $\mu\rightarrow \infty$ with $q=1$, then the butterfly speed does go to zero since since spin flips in this fully polarized background can move at most diffusively. Thus, the $q=1$ problem as $\mu\rightarrow \infty$ is closer to the zero temperature dynamics of Hamiltonian systems. In the rest of this section, we will work at $q=1$ in the $\mu \rightarrow \infty$ limit. 

We start with $\mu=\infty$ where $\rho_\infty^{\rm eq}$ reduces to the projector on the $|0\rangle \equiv |\uparrow \uparrow \cdots \uparrow\rangle$ state.  Then, 
\begin{equation}
\mc{C}^\infty_{\alpha\beta} = \frac{1}{2} \langle 0| |[\sigma_0^\alpha(t), \sigma_x^\beta]|^2|  0 \rangle. 
\end{equation}
First consider OTOC's involving $z$ and $r$. Since $|0\rangle$ is an eigenstate of both $r$ and $z$ (with eigenvalues 0 and 1 respectively), any commutator involving only $r$'s or $z$'s must be strictly zero. Thus, the only non-trivial OTOCs are those involving $[l_0(t), z_x]$ and $[l_0(t), r_x]$ (the commutators between $[z_0(t), l_x]$  and $[r_0(t), l_x]$ will be related to these by time-reversal). Let us start with  $\mc{C}^\infty_{lz}$. Expanding the commutator, we find: 
\begin{align}
\mc{C}^\infty_{lz} &= \frac{1}{2} \langle 0| |[l_0(t), z_x]|^2|  0 \rangle \nonumber \\
&=  \frac{1}{2} \langle 0| |\left(z_x r_0(t) -  r_0(t)z_x\right)\left(l_0(t)z_x - z_x l_0(t)\right)|  0 \rangle \nonumber \\
&=  2 -  \langle 0|  r_0(t) z_x l_0(t)|  0 \rangle
\label{eq:infinitemuotoc}
\end{align}
where we've made use of the fact that $|0\rangle$ is an eigenstate of $U(t), z$ and  $r_0l_0$ in the last line, with eigenvalues 1,1, and 2 respectively. Now, every basis string that appears in the operator expansion of $l_0(t)$ has net raising charge $\mc{R}^{\rm tot} = -1$. Since $|0\rangle$ is annihilated by any $r_i$, all strings in the expansion of $l_0(t)$ that contain $r$'s cannot contribute to the OTOC, and thus only strings containing exactly one $l_i$ on some site (with arbitrary $\I$'s and $z$'s on the others) can contribute. These create single spin flips in the polarized background. 
Of these, only the strings which contain $l_x$ fail to commute with $z_x$. 
However, due to the diffusion of ``raising charge'' $\mc{R}_i$ discussed in Section~\ref{sec:raisingandspin}, the weight of the operator on such ``single $l$" strings decays only diffusively away from the origin, showing that this OTOC displays only diffusive (as opposed to ballistic) behavior corresponding to a zero butterfly speed. Likewise, consider the infinite $\mu$ OTOC involving $[l_0(t), r_x]$:
\begin{align}
\mc{C}^\infty_{lr} &= \frac{1}{2} \langle 0| |[l_0(t), r_x]|^2|  0 \rangle \nonumber \\
&= \frac{1}{2} \langle 0|  r_0(t) (l_xr_x) l_0(t)|  0 \rangle\nonumber \\
 &= \frac{1}{2} -\frac{1}{2}\langle 0|  r_0(t) z_x l_0(t)|  0 \rangle,
\end{align}
where we've made use of the fact that $r_x|0\rangle = \langle 0 | l_x = 0$, and $(lr) = (1-z)$. Thus, this OTOC has an identical structure to  \eqref{eq:infinitemuotoc} and also displays purely diffusive behavior. Perturbing $\rho_\mu^\infty$ away from $\mu = \infty$ to leading order in large but not infinite $\mu$ gives~\eqref{eq:rhomu}
\begin{equation}
\rho_{\mu\rightarrow \infty}^{\rm eq}\approx (2)^{-L} \prod_i \sqrt{2} \left[(1-e^{-2\mu}) u_i  + e^{-2\mu} d_i\right],
\end{equation}
Understanding the modifications to the diffusive $\mu = \infty$ behavior of the OTOCs at large but finite $\mu$ is an interesting question that has been explored in detail in Ref.~\onlinecite{TiborCons}. 

\section{Physical systems}
\label{sec:physical}
Let us now turn to more generic examples of thermalizing spin chains with conservation laws, and examine to what extent the universal aspects of our results continue to hold in this setting. Since analytic results for the dynamics are not easily obtained for these systems, we will study them numerically using exact diagonalization. 
We will first look at a periodically driven ``Floquet'' spin chain where energy is not conserved, but $S_z^{\rm tot}$ is, just like in our random circuit model. We then turn to a interacting quantum chaotic spin chain with energy conservation. 

\subsection{$\Sz$ Conserving Floquet chain}
We consider a one-dimensional chain of spin-1/2 qubits that is periodically driven in time with period $\tau$ between two separate  Hamiltonians $H_z$ and $H_{xy}$, each of which act for half the period. The time evolution operator for this system is
\begin{align}
U_F(\tau) &= e^{i \frac{\tau}{2} H_z}e^{i \frac{\tau}{2} H_{xy}}, \nonumber \\
\end{align}
with 
\begin{align}
H_z &= J_z \sum_i \sigma_i^z \sigma_{\ip}^z,  \nonumber \\
H_{xy} &= \sum_i  [J_x \sigma_i^x \sigma_{\ip}^x + J_y \sigma_i^y \sigma_{\ip}^y]~.
\label{eq:UFloq}
\end{align}
The parameters we choose for this model are $J_z = (10+\sqrt{5})/16$, $J_x = J_y = (5+5\sqrt{5})/16$, and $\tau = 1$, where we have found numerically that this system is well thermalizing, even though the individual Hamiltonians in the drive are integrable\footnote{We choose irrational numbers $J_x, J_z$ that are order 1, since this choice has been shown to thermalize better for small system sizes compared to choosing rational parameters\cite{KimHuse}}.  The time-evolution conserves total $\Sz$: $[U(\tau), \Sz]=0$ and, just as in the random circuit problem, operators like $r/l$ have a definite $\Delta\Sz$ action which is respected by the time evolution.  We will measure time discretely in multiples of $\tau$, and the time evolution of operators in the Heisenberg picture is given by $O(n\tau) = U_F^\dagger(n\tau) O(0) U_F(n\tau)$.  Finally, there is no randomness in this problem, and the time evolution respects locality. 

Of course, the fact that charges in this thermalizing model will display diffusive dynamics is well understood. We wish to probe the interaction between the diffusive charge dynamics and the ballistic operator growth, say as probed through the right-weight profiles $\rho_R(x,t)$ of $r_0(t)$ and $z_0(t)$. While $\rho_R$ for $z(t)$ does show a tail for this model, it is difficult to tease apart the different regimes for small finite-sized systems accessible to exact-diagonalization studies (it only takes time $t\sim L^2$ for the diffusive charges to reach the end of the chain). To aid with this, we consider a related quantity, the $\ell$-weight $W_\ell(x,t)$ which measures the weight on all Pauli strings with maximum end-to-end separation between non-identity elements equal to $x$:
\begin{align}
W_\ell(x,t) = \sum_{\substack{{\text{ $\mc{S}:$ LHS$(\mc{S})-$} }  \\  {\text{  RHS$(\mc{S}) = x$ }} }} |a_{\mc S}|^2
\label{eq:lweight}
\end{align}
where, as before, the LHS and RHS of $\mc{S}$ define the locations of the rightmost and leftmost non-identity operators in $\mc{S}$. This has the advantage that all conserved charges have the form $z_i$  and these are all mapped to $x=0$. On the other hand, strings that are part of the ballistic front and stretched between $-\vb t,\vb t$ contribute to $x=2\vb t$. Thus, the non-conserved contributions to this quantity still show ballistic dynamics. In a system with no conservation laws, $W_\ell(x,t)$ for  $x< \vb t$ decays exponentially with $t$ and quickly approaches its asymptotic value which is itself exponentially small in $L$. On the other hand, in a system with conservation laws,  this quantity decays as a power law and approaches its asymptotic value which is instead only power law small in $L$ if one starts with an initial operator which has overlap with the conserved densities (at late times, each conserved amplitude $\sim 1/L$ so that the total \emph{weight} on all conserved charges is $\sim L/L^2 = 1/L$). This helps us tease apart the dynamics of the conserved charge at these small sizes. 

\begin{figure}
  \includegraphics[width=\columnwidth]{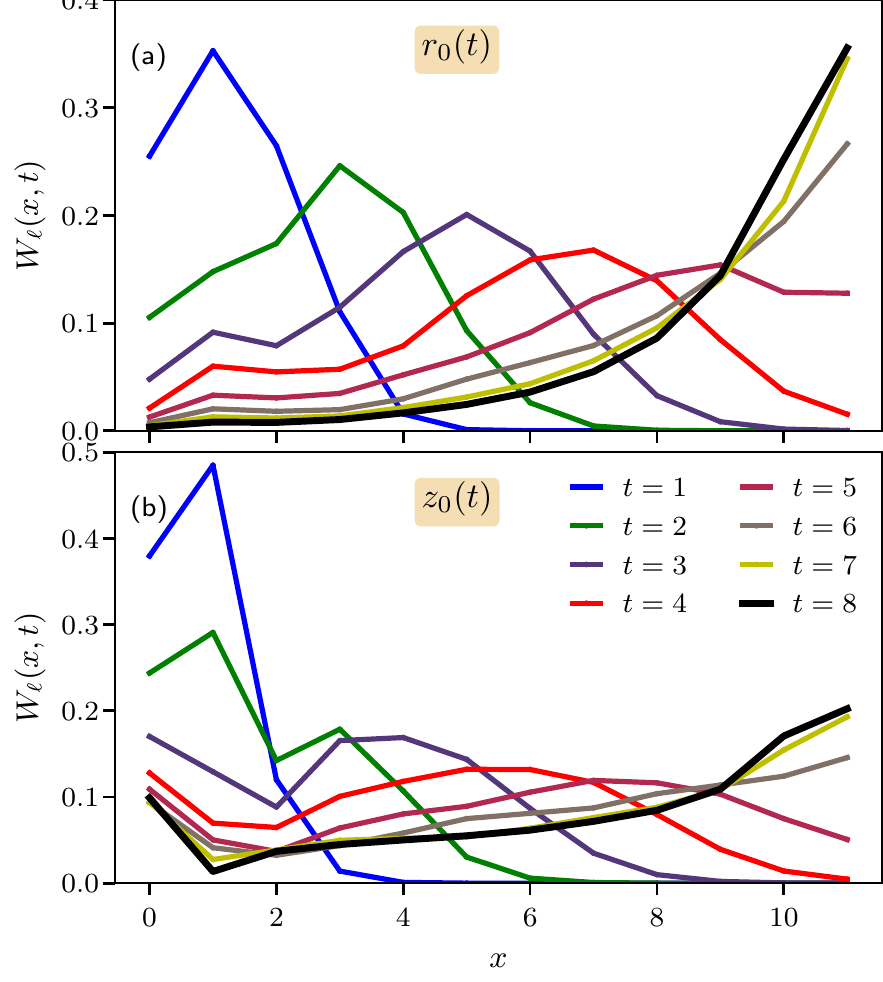}
  \caption{\label{fig:floqshape} $\ell$-weight profile $W_\ell (x,t)$ \eqref{eq:lweight} of spreading operators $z_0(t)$ and $r_0(t)$ in a Floquet model that conserves $\Sz$ \eqref{eq:UFloq},  plotted for a system of length $L=12$. The expected diffusive ``lump'' in the right-weight profile of the spreading conserved charge is manifested in the enhanced weight of $W_\ell (x,t)$ at  $x=0$ for $z_0(t)$, even at late times (b). No such enhancement is observed for $r_0(t)$ which evolves orthogonal to the conserved charges (a).  Notice also that the late-time decay of $W_\ell (x,t)$ away from $x=L$ is much slower for $z_0(t)$  as compared to $r_0(t)$, consistent with the presence of power law diffusive tails in the right-weight profile of the former but not the latter.       }
\end{figure}

Fig.~\ref{fig:floqshape} shows the $\ell$-weight profiles $W_\ell(x,t)$ for both $r_0(t)$ and $z_0(t)$ for a 12 site chain, with both operators starting at the end of the chain to maximize the available distance for spreading.  We clearly see the tail and ``lump'' with enhanced $\ell$-weight at $x=0$ for $z_0(t)$. On the other hand,  the late time profile for $r_0(t)$ shows a simple decay of the $\ell$-weight, with the most weight on strings stretched across the entire system \emph{i.e.} those with $x=L$. This data shows that the central aspects of our results on operator shape seem to be born out for this more ``physical'' spin chain, although for such a short chain this can only be qualitatively tested in the numerics. 

\subsection{Hamiltonian spin chain}

\begin{figure}
  \includegraphics[width=\columnwidth]{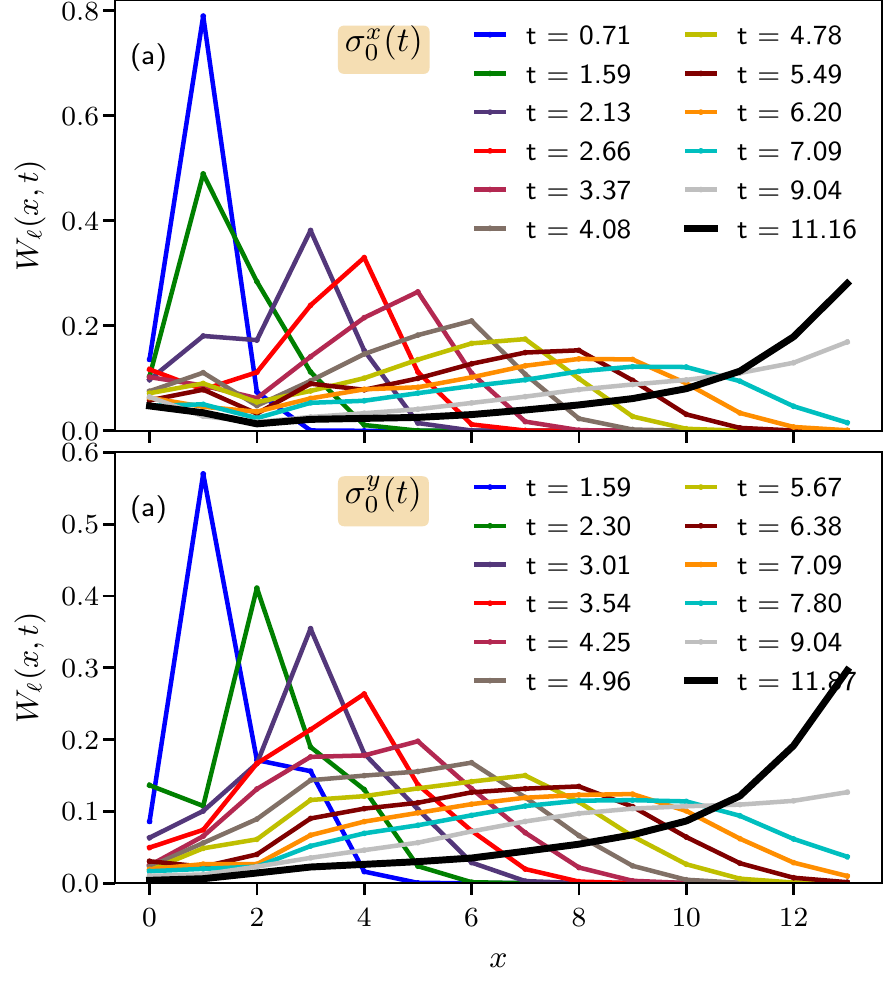}
  \caption{\label{fig:KHshape} $\ell$-weight profile $W_\ell (x,t)$ \eqref{eq:lweight} of spreading operators $\sigma^x_0(t)$ and $\sigma^y_0(t)$ in a thermalizing Hamiltonian model that conserves energy \eqref{eq:KimHuse},  plotted for a system of length $L=14$. $\sigma^x_0$ has non-trivial overlap with the local conserved energy density operators, which is visible in the enhanced weight of $W_\ell(x,t)$ for $x=0,1$ even at late times (a). By contrast, $\sigma^y_0$ starts out orthogonal to the conserved charges and its late-time $\ell$-weight profile does not show a ``lump'' at $x=0,1$. While the late time decay of $W_\ell (x,t)$ away from $x=L$ is again faster for $\sigma^y_0(t)$ as compared to $\sigma^x_0(t)$, the difference is not as pronounced as the difference between $z_0(t)$ and $r_0(t)$ in the $U(1)$ Floquet case. This is because $\sigma^y_0(t)$ can develop overlap with the slow conserved charges, unlike $r_0(t)$ which is constrained to remain strictly orthogonal to the charges during its entire evoluton. }
\end{figure}

We now turn to a thermalizing Hamiltonian spin-chain with energy conservation, but no other continuous symmetries\cite{KimHuse}:
\begin{align}
H = \sum_i J \sigma_i^z \sigma_{\ip}^z + h_x \sigma_i^x + h_z \sigma_i^z,
\label{eq:KimHuse}
\end{align}
where $J = 1, h_x = (\sqrt{5}+5)/8, h_z = (\sqrt{5}+1)/4$. The local energy density operators are one and two-site Pauli strings of the form $\sigma^x_i$, $\sigma^z_i$ and $\sigma^z_i \sigma^z_{\ip}$. The conservation of total energy implies that $\mbox{Tr}\; (O_0(t) H) = \mbox{const}$, and thus if we start with an operator with a non-zero overlap with a local energy density, then the sum of the amplitudes of the spreading operator on all the local energy density strings is conserved. A difference between this Hamiltonian model and the $U(1)$ conserving model is that energy is not quantized and there are no special operators (like $r$) with definite algebras under $H$. Even if we start with a local operator that is orthogonal to all the conserved energy densities (like $\sigma^y_i$), its operator expansion will, in time, develop some overlaps with the conserved charges.

Fig.~\ref{fig:KHshape} shows the $\ell$-weight profile for the spreading of $\sigma^x_0(t)$ and $\sigma^y_0(t)$ in a system of length $L=14$ obtained using exact-diagonalization and, once again, we see the ``lump'' at $x=0,1$ only in the former. Notice, however, that the difference between $\sigma^x_0$ and $\sigma^y_0$ in the decay of $W_\ell(x,t)$ away from $x=L$ at late times is not as pronounced as it is in the Floquet problem (Fig.~\ref{fig:floqshape}). We attribute this to the fact that the time-evolving $\sigma^y_0(t)$ develops non-trivial overlap with the diffusing conserved charges (even though the sum of the amplitudes on all conserved charges is constrained to be zero), slowing down its dynamics relative to $r_0(t)$ in the Floquet model which always evolves strictly orthogonal to the conserved charges. This is one example of a difference between $U(1)$ conservation with quantized charges and energy conservation where the charges are continuous.  

As discussed in the previous section,  these power law tails in operator shape translate into power law tails in the long distance and late time OTOC, and these have been observed numerically in Hamiltonian systems with conservation laws\cite{FradkinHuse}. 

\section{Conclusions}
\label{sec:conclusion}
In conclusion, we presented an extensive study of the ``scrambling'' dynamics of local operators in chaotic quantum systems with a conserved, diffusing charge (or energy) density.
A generic local operator in this setting has some weight on the conserved charges, and the late time spreading dynamics of such an operator is described by multiple coupled hydrodynamic processes. The first is the ``physical'' hydrodynamics associated with the diffusive dynamics of the conserved charges.  We show that the total operator weight on these conserved charges decreases as power law in time which, by unitarity, necessitates a steady ``emission'' process that transfers operator weight from local conserved densities to nonconserved operators. This emission happens at a slow hydrodynamic rate set by the local diffusive currents of the conserved density but, once emitted, the fronts of the nonconserved operators spread ballistically and rapidly become nonlocal.  The propagation of the nonconserved fronts is described via an ``emergent hydrodynamics'' that is biased diffusion for the one-dimensional case\cite{opspreadAdam,opspreadCurt}, and this coupled diffusion-emission-propagation process reveals a composite picture for the operator profile, showing how the ballistic and diffusive processes in the system connect at different time and length scales.  In particular,  the presence of slow diffusive modes leads to the development of a power-law tail in the operator profile that reflects the weight of nonconserved operator strings emitted at later times that ``lag'' behind the leading ballistic front. 

Our picture illustrates how reversible unitary dynamics in closed quantum systems can display dissipative diffusive hydrodynamic modes.  The dissipation arises from the conversion of operator weight from locally observable conserved parts to nonlocal nonconserved parts at a slow hydrodynamic rate, a process which increases the ``observable" entropy of the system.  By contrast, in systems without conservation laws, any local operator is rapidly converted to non-local, so the dissipation does not appear in the late-time hydrodynamics of operator or entanglement spreading.    

In addition to the diffusive tails in the distribution of operator weight, we found an additional layer of structure describing the local operator content behind the ballistic front, {\it within} the spreading operator.  Outside the ballistic operator front, the spreading operator locally consists only of local identities.  The arrival of the front at a given site turns on a noisy coupled diffusion process between different species of local operators, which relaxes the initially imbalanced local operator content to the final equilibrium value where all local operators are equally likely.  Once again, this relaxation happens at a power-law slow rate and contributes to the operator hydrodynamics, in contrast to the unconstrained random circuit case where the action of a single gate at the front erases the initial bias towards the identities.  These power law tails in the distributions of operator weight and local operator content also lead to diffusive tails in the late-time approach of certain out-of-time-order commutators (OTOC's) to their asymptotic values. 

In all, our work reveals several rich layers of physics in the scrambling dynamics of systems with conservation laws.  We expect this approach (which builds on work by Refs~\onlinecite{AdamCircuit1, opspreadAdam, opspreadCurt}) of probing the dynamics of such systems via an analytically tractable constrained random unitary circuits will have broader applicability in understanding the many open questions about the fundamentals of thermalization and quantum statistical mechanics in chaotic quantum systems with conservation laws.  It would also be interesting to ask which of our results can also be calculated (or extended) in a holographic setting, where they might connect with the dynamics of black holes with charges. 

\vspace{2pt}
\emph{Related Work:} Shortly before completing this manuscript, we became aware of related work by Rakovszky \emph{et. al.} which should appear in the same arXiv posting\cite{TiborCons}.  While these authors take a somewhat different approach, our results appear to agree where they overlap.  

\section*{Acknowledgements}
We thank Cheryne Jonay, Joel Lebowitz, Adam Nahum, Stephen Shenker, Shivaji Sondhi, Douglas Stanford and Brian Swingle for helpful discussions. VK was supported by the Harvard Society of Fellows and the William F. Milton Fund.  AV was supported by a Simons Investigator grant. 

\bibliography{global}

\begin{appendix}

\section{Action of $S_z^{\rm tot}$ conserving unitary on different operators}
\label{sec:circuitaction}
There are a few classes of operators on site $i$ that evolve differently under the action of $U$. These are: $(r a)_i$, $(l a)_i$, $(z \I)_i$, $(zn)_i$, $(\I \I)_i$, and $(\I n)_i$.  
We  use the labels $a = \mbox{``any''}$ and $n = \mbox{``non-identity''}$ to refer to the set of qudit operators that take ``any'' value or that act as the ``non-identity'' respectively. 
If one starts with an operator with a definite action (such as ``raise by one'') under the U(1) symmetry, the circuit preserves this action.  The action of $U_{i, i+1}$ on two-site operators  can be summarized as follows (in all cases the norm of the operator is preserved):
\begin{enumerate}
\item The operators $(\I\I)_i(\I\I)_{i+1}$, $(z \I)_i(z \I)_{i+1}$,  and $[(z\I)_i(\I\I)_{i+1}~+~(\I\I)_i(z\I)_{i+1}]/\sqrt{2}$ are invariant under the action of the gate due to the conservation law. 

\item The $4q^4$ raise-by-one operators  $(r a)_i(\I a)_{i+1}$, $(r a)_i(z a)_{i+1}$, $(\I a)_i(r a)_{i+1}$ and  $(z a)_i(r a)_{i+1}$ each transition to a uniformly random linear combination of these operators.  Similarly, the lower-by-one operators  $(l a)_i(\I a)_{i+1}$, $(l a)_i(z a)_{i+1}$, $(\I a)_i(l a)_{i+1}$ and  $(z a)_i(l a)_{i+1}$ each transition to a uniformly random linear comination of these $4q^4$ operators.  

\item  The $q^4$ raise-by-two operators  $(r a)_i(r a)_{i+1}$ each transition to a uniformly random linear combination of these operators.  Likewise, the lower-by-two operators  $(l a)_i(l a)_{i+1}$ each transition to uniformly a random linear combination of these $q^4$ operators.

\item The $(6q^4-3) $ $\Sz$-conserving operators $(r a)_i(l a)_{i+1}$, $(l a)_i(r a)_{i+1}$, $(\I a)_i(\I a)_{i+1}$, $(z a)_i(z a)_{i+1}$, $(\I a)_i(z a)_{i+1}$ and $(z a)_i(\I a)_{i+1}$ (excluding the three special operators in case 1 above) each transition to a random linear combination of these operators, but here with non-uniform probabilities.  This is due to the fact that the $\I$'s and $z$'s are superpositions of the conserved `up' and `down' spin charges introduced in Section~\ref{sec:raisingandspin}. 

Thus, it is more convenient to express these transitions directly in in terms of the up and down qubit projection operators $u= \frac{(\I+z)}{\sqrt{2}}$ and $d= \frac{(\I-z)}{\sqrt{2}}$:  
\begin{enumerate}
\item $(u\I)_i(u\I)_{i+1}$, $(d\I)_i(d\I)_{i+1}$ and $[(u\I)_i(d\I)_{i+1}~+~(d\I)_i(u\I)_{i+1}]/\sqrt{2}$ are left invariant by the action of the circuit. 

\item The $(q^4-1)$ operators $(u a)_i(u a)_{i+1}$ [but not including the conserved $(u\I)_i(u\I)_{i+1}$] each transition to a uniformly random linear combination of these operators.  
\item Likewise, the $(q^4-1)$ operators $(d a)_i(d a)_{i+1}$ [but not including the conserved $(d\I)_i(d\I)_{i+1}$] each transition to a uniformly random linear combination of these operators.  
\item The $(4q^4-1)$ operators $(r a)_i(l a)_{i+1}$, $(l a)_i(r a)_{i+1}$, $(u a)_i(d a)_{i+1}$ and $(d a)_i(u a)_{i+1}$ [not including the conserved $[(u\I)_i(d\I)_{i+1}~+~(d\I)_i(u\I)_{i+1}]/\sqrt{2}$] each transition to a uniformly random linear combination of these operators.
\end{enumerate}
\end{enumerate}

Finally, for completeness, we note that the raise/lower-by-one operators in case two above split into two groups each in the $u,d$ basis:
\begin{enumerate}[(a)]
\item The $2q^4$ operators of the form  $(r a)_i(u a)_{i+1}$, $(u a)_i(r a)_{i+1}$ mix between each other with equal probability $1/2q^4$. These act on states (on this two site subspace) with $\Sz=0$ and produce states with $\Sz = 1$. 

\item The $2q^4$ operators of the form  $(r a)_i(d a)_{i+1}$, $(d a)_i(r a)_{i+1}$ mix between each other with equal probability $1/2q^4$. These act on states (on this two site subspace) with $\Sz=-2$ and produce states with $\Sz =-1$. 

\item The $2q^4$ operators of the form  $(l a)_i(u a)_{i+1}$, $(u a)_i(l a)_{i+1}$ mix between each other with equal probability $1/2q^4$. These act on states (on this two site subspace) with $\Sz=2$ and produce states with $\Sz = 1$. 

\item The $2q^4$ operators of the form  $(l a)_i(d a)_{i+1}$, $(d a)_i(l a)_{i+1}$ mix between each other with equal probability $1/2q^4$. These act on states (on this two site subspace) with $\Sz=0$ and produce states with $\Sz = -1$. 
\end{enumerate}

Note that in the $u,d,r,l$ basis, the action of the circuit looks symmetric with respect to interchanging $r \leftrightarrow l$ and $u \leftrightarrow d$. 

\section{Derivation of the butterfly speed $v_B$ and the diffusion constant $D_\rho$ to $O(1/q^2)$}
\label{sec:vb}
In this section we derive the butterfly speed $\vb$ and the diffusion constant $D_\rho$ characterizing the biased diffusion of the nonconserved operator fronts.  We assume we are working at late times so the ballistic front of nonconserved operators is well separated from the diffusive lump of conserved charges near the origin. We present the discussion below for the propagation of the right-front, but the same holds for the left front as well, just reflected. 

We find it convenient to define the right front of an operator string as the location of the rightmost two-site unitary \emph{gate} that sees a non-identity operator at time $t$. So, for instance, if an operator string ends on site $i$ at time $t$, the front (for this string) is either between sites ${(i, i+1)}$ or between sites ${(i-1,i)}$, depending on the evenness and oddness of $t,i$ due to the even-odd staggered structure of the circuit (Fig.~\ref{fig:circuit}). Without loss of generality, let's consider a particular operator string for which the front at time $t$ is at the gate on sites ${(i, i+1)}$. Now, under the action of the circuit, the front gate has a probability that is suppressed as $O(1/q^2)$ for creating $(\I\I)_{i+1}$ on site $(i+1)$ (Appendix~\ref{sec:circuitaction}) (unless the string's local operators at the front gate are $(z\I)_i(\I\I)_{\ip}$, $(\I\I)_i(z\I)_{\ip}$ or $(z\I)_i(z\I)_{\ip}$; we return to these cases below which are themselves suppressed in probability by $O(1/q^2)$, $O(1/q^4)$ and $O(1/q^4)$, respectively).
When the front gate creates an identity on site $(i+1)$, the string ends on site $i$ at time $(t+1)$. But, due to the even-odd staggering of gates, this means that the front at time $(t+1)$ is on the sites $(i-1, i)$ which means it moved \emph{back} one step. On the other hand, if the front gate at time $t$ produces a non-identity on site $(\ip)$ (which happens with probability $1- O(1/q^2)$), then the front at time $(t+1)$ is on sites $(\ip, i+2)$, which means it moves forwards one step.  At $q=\infty$, the front deterministically moves forwards at every time step which gives a butterfly velocity $\vb = 1$.  Also, in this limit, the front of the full initially local operator remains perfectly sharp since there is no probabilistic spread in the locations of the fronts of the strings that make up the resulting spreading operator. 

We now estimate the leading $1/q^2$ correction to the butterfly speed.  To do this, we need to estimate the probability of the front moving backwards.  Note that the most probable configuration of local operators on the front gate takes the form $(n)_i (\I\I)_{i+1}$ where $n$ is a non-identity.  This is because if the action of the front gate $(i-1,i)$ at time $(t-1)$ produces a non-identity on site $i$ (which is the most probable option at large $q$), then this non-identity meets the identity on site $(i+1)$ on the front gate at the next time step which is at $(i, i+1)$. The different types of front operators that belong to this category are: (i) $(ra)_i (\I\I)_{i+1}$, (ii) $(la)_i (\I\I)_{i+1}$, (iii) $(\I n)_i (\I\I)_{i+1}$, (iv) $(zn)_i (\I\I)_{i+1}$, and (v) $(z\I)_i (\I\I)_{i+1}$, where we've split the $z$ case into two for reasons that will become clear momentarily. To estimate the probability of the front moving backwards, we need to estimate the steady-state probability of getting these different operator types at the front, and the probability of the different types moving back. We will find that cases $(i) - (iv)$ exist with $O(1)$ probabilities and move back with probability $O(1/q^2)$, while case $(v)$ exists with probability $(1/q^2)$ and moves back with probability $O(1)$. Thus, each of these gives an $O(1/q^2)$ probability of moving backwards. Note that we won't need to consider front operators of the form $(a)_i (n)_{i+1}$ at this order, because only way to produce front operators of this form is for the front to have moved backwards at the previous step $(t-1)$ \emph{i.e.} the front at $(t-1)$ was at sites $(i+1, i+2)$ and created an identity on site $(i+2)$ leading to the front moving to sites $(i,i+1)$ at time $t$, where the non-identity on site $(i+1)$ now generically meets a non-identity on the ``backward'' site $i$ with probability $O(1-1/q^2)$. Thus, the front gate sees motifs like $(n)_i (n)_{i+1}$ with probability  $O(1/q^2)$ and they move back with probability $O(1/q^2)$, thereby giving a higher order $O(1/q^4)$ correction to $\vb$. The case where the front sees $(\I\I)_i(z\I)_{\ip}$ is further reduced by a factor of $1/q^2$ since this case both requires the front to move back at the previous time step, and for the qudit operator on site $i$ to be the identity. 

We now turn to the steady-state probabilities of the different operator types (i)-(v) at the front. Since we only need the probabilities for $(i)- (iv)$ correct to $O(1)$, we can estimate them at infinite $q$. 
The infinite $q$ answer for $\Delta_{rs}$ \eqref{eq:deltars_exact} evaluated at the location of the front $x=t$, combined with the fact that $\rho_r=\rho_l$ and $\rho_u = \rho_d$ at the front tells us that, at infinite $q$, the front gate looks like $(ra)_i (\I\I)_{i+1}$ or $(la)_i (\I\I)_{i+1}$  with equal probability $1/6$, and it looks like $(un)_i (\I\I)_{i+1}$ or $(dn)_i (\I\I)_{i+1}$ with probability $1/3$. The latter two cases translate to an equal $1/3$ probability for obtaining  $(zn)_i (\I\I)_{i+1}$ and  $(\I n)_i (\I\I)_{i+1}$ at the front gate. On the other hand, the probability for seeing a $(z\I)_i (\I\I)_{i+1}$ at the gate is only $O(1/q^2)$. This is because, away from the location of the diffusing charges,  processes that generate such an operator at a site have $q^2$ choices for the operator on the qudit spin, and an $\I$ on the qudit spin reflects only one of these $q^2$ choices. Thus, we denote the probability of obtaining $(z\I)_i (\I\I)_{i+1}$ at the front as  $(p_z/q^2)$, and self consistently solve for $p_z$ below. 

\begin{enumerate}
\item The input to the last gate is $(ra)_i (\I\I)_{i+1}$ or $(la)_i (\I\I)_{i+1}$ with equal probability $1/6+O(1/q^2)$. These produce a $(\I\I)_{\ip}$ on site $(i+1)$ with probability $1/(4q^2)$ (Appendix \ref{sec:circuitaction}) which makes the front move back. Or they produce ($(z\I)_{\ip}$) on site $(\ip)$ with probability $1/(4q^2)$,which results in the front gate at the time $(t+1)$ seeing the operator $(z\I)_i (\I\I)_{i+1}$. 

\item With probability  $2/3 + O(1/q^2)$  the last gate gets input  $(zn)_i (\I\I)_{i+1}$ or  $(\I n)_i (\I\I)_{i+1}$  
These are either $(un)(u\I) , (dn)(d\I)$, $(un)(d\I)$ or $(dn)(u\I)$ with equal probabilities, where we've expressed the qubit operators $z$ and $\I$ as an superpositions of $u$ and $d$.  Now, under the action of the circuit,  $(un)(u\I) $ transforms into a uniformly random linear combination of all $(ua)(ua)$ except $(u\I)(u\I)$, and  $(dn)(d\I)$ does likewise (Appendix~\ref{sec:circuitaction}).  Thus, these only make $z$ and $\I$ qubit operators at these two sites.  On the other hand,  $(un)(d\I)$, $(dn)(u\I)$ equally make all four qubit operator types $(r,l,z,\I)$, since these mix between $(r a)_i(l a)_{i+1}$, $(l a)_i(r a)_{i+1}$, $(u a)_i(d a)_{i+1}$ and $(d a)_i(u a)_{i+1}$ with equal probability to order $1/q^2$.  As a result, given this input, the outputs on qubits are $\I$'s with probability $3/8$, $z$'s with probability $3/8$, and otherwise $r$'€™s or $l$'s.  So for these cases the resulting output on the front site $(i+1)$ is $(\I\I)_{\ip}$ with probability  $3/(8q^2)$  and this results in the front moving backwards.  Similarly, the resulting output on the front site is $(z\I)_{\ip}$ with probability $3/(8q^2)$, which results in the front gate at the time $(t+1)$ seeing the operator $(z\I)_i (\I\I)_{i+1}$. 

\item  Finally, with probability  $p_z/q^2$  the last gate gets input $(z\I)_i (\I\I)_{i+1}$  (where $p_z/q^2$ is a steady state probability yet to be determined).  With probability $1/4 + O(1/q^2)$ this is left alone, so the front moves back.  Or the resulting output on the front site is $(z\I)_{\ip}$ with probability $(1/4+ O(1/q^2)$. This follows since $(z\I)_i (\I\I)_{i+1} = \frac{(z\I)_i (\I\I)_{i+1} + (\I\I)_{i}(z\I)_{\ip} }{2} + \frac{(z\I)_i (\I\I)_{i+1} - (\I\I)_{i}(z\I)_{\ip} }{2}$, and the former is conserved under the action of the circuit while the latter mixes between $q^4$ different operators. The conservation of the former gives the leading $1/4$ probability for producing $(z\I)_i$ or $(z\I)_{\ip}$. 

\end{enumerate} 

We can now use (i)-(iii) to self-consistently solve for the steady state value of $p_z$ at the front:
$$\frac{p_z}{q^2}(t+1) = \frac{1}{4q^2} \frac{2}{6} +   \frac{3}{8q^2} \frac{2}{3} + \frac{1}{4} \frac{p_z}{q^2}(t) = \frac{p_z}{q^2}(t)$$
where the last equality is true in the steady state. Solving this gives $p_z=4/9$.  Finally  putting  together the probability of each operator type at the front with probability for a given type to move back   gives a net probability for the front to move back equal to $p_{\rm backwards} = 4/9q^2$, which means the butterfly velocity is $\vb = (p_{\rm forward} - p_{\rm backwards}) = 1-8/(9q^2)$.  

Next, we estimate the diffusion constant for the biased random walk of the operator front. After $t$ steps, the expectation value of the location of the front is denoted $\langle X \rangle _t = \langle \sum_{i=1}^{t} x_i \rangle $ where $x_i = -1 $ with probability $p_{\rm backwards}$ and $x_i = +1$ with probability $1- p_{\rm backwards}$. Then, 
\begin{align*}
\Delta_X &\equiv \left \langle \left(\sum_{i=1}^t x_i\right )^2 \right \rangle - \left \langle \sum_{i=1}^t x_i \right \rangle^2 \\
&= t +  \left \langle \sum_{i\neq j} x_i x_j  \right \rangle  - (\vb t)^2. 
\end{align*}
Unlike a Markovian random walk where $x_i$ and $x_j$ at different time steps are independent, the constrained action of the circuit does introduce correlations between $x_i$ and $x_j$. However, these only appear at $O(1/q^4)$ and higher. To see this, one can look at the modified probabilities of the the front operators at time $t$ \emph{given} that the front moved forwards/backwards at time $t-1$ (and other such processes), and we find that all of these only receive higher order corrections at $1/q^4$. Thus, to leading order in $1/q^2$, the random walk does look Markovian which gives a diffusion constant equal to $D_\rho = (1-\vb^2)/2 \approx 8/(9q^2)$. 

\section{Amplitudes and weights of the evolving operator on the conserved densities}
\label{sec:consamps}

To see the evolution of the conserved amplitudes $a_i^c(t)$ under the unitary dynamics, consider the action of the gate $U_{12}$ on $(a_1^c(z\I)_1 + a_2^c(z\I)_2)$:
\begin{align}
a_1^c(z\I)_1 + a_2^c(z\I)_2 &= (a_1^c+a_2^c) \left[\frac{(z\I)_1(\I\I)_2+ (\I\I)_1(z\I)_2}{2}\right]\nonumber\\
&+(a_1^c-a_2^c)\left[\frac{(z\I)_1(\I\I)_2- (\I\I)_1(z\I)_2}{2}\right].
\label{eq:action_a}
\end{align}
The first term above is unchanged under the action of $U_{12}$, and the second term is  a non-conserved operator  since its net overlap with $S_z^{\rm tot}$ is zero. Under the action of $U_{12}$, the second term transitions to a random superposition of of order $q^4$ operators that locally conserve $S_z^{\rm tot}$ but act arbitrarily on the qudit (Appendix~\ref{sec:circuitaction}). 
Only one of these operators --- corresponding to the second term returning to itself  --- contributes to the amplitudes of the conserved charges. The amplitude for this term is of order $1/q^2$, but it is random and vanishes upon averaging over different random circuits.  Thus, after the action of the gate, the Haar-averaged amplitude of the conserved local operator on both outgoing sites is equal to the average of the incoming amplitudes:
\begin{equation}
\overline{a^c_1(t+1)} = \overline{a^c_2(t+1)}= \frac{\overline{a_1(t)}+\overline{a_2(t)}}{2}.
\label{eq:acircuit_app}
\end{equation}
Starting with $(z\I)_0$ on a single site and recursively applying this formula gives a ``doubled" version of Pascal's triangle (due to the even-odd structure of the circuit) according to which the amplitudes on the different sites are: 
\begin{align}
\overline{a_i^c(t)} = \frac{1}{2^t}   {{t-1}\choose{\floor{\frac{i+t-1}{2}}}}.
\end{align}

Additionally, as a result of the ``equalizing" action \eqref{eq:acircuit_app}, the circuit-to-circuit variance in the conserved amplitudes is suppressed both in the  large $q$ and the late $t$ limit:
\begin{equation}
\Delta_i^{a}(t) \equiv \overline{|a_1^c(t)|^2}  - |\overline{a_1^c(t)}|^2  \sim \frac{1}{q^4}\frac{1}{t^2},
\label{eq:anoise2_app}
\end{equation}
This can be  understood from \eqref{eq:action_a} since the difference $\Delta_1^a(t)$ is proportional to the weight of non-conserved terms $\sim|(a^c_1(t) - a^c_2)|^2 \sim (\partial_x a^c_x)^2$ which scales as  $\sim 1/t^2$ in the region $x \sim \sqrt{t}$ where $a_i^c(t)$ is appreciable \eqref{eq:adiff}. This suppression reflects the ``smoothing'' action of the circuit which locally makes the averaged amplitudes of the conserved charges equal and thus reduces $(\partial_x a^c_x)$ in time.  Further, the additional ($1/q^4$) suppression comes from the fact that only transitions of non-conserved operators to $[( z\I)_i - (z\I)_{\ip})]$ contribute to $\Delta_i^a(t)$, and this transition is only one of of order $q^4$ possibilities. Likewise, we can see how this process introduces correlations in the amplitudes $a^c_i$ and $a^c_{\ip}$ and 
\begin{equation}
 \overline{|a_1^c(t) a_2^c (t)|}  -  \overline{|a_1^c (t)|}\;\overline{| a_2^c (t)|}  \sim \frac{1}{q^4}\frac{1}{t^2}.
\label{eq:anoise3}
\end{equation}

\section{Two-point correlations of spin and raising charge}
\label{sec:Cij}
We derive \eqref{eq:deltars_exact} at $q=\infty$.  At infinite $q$, the inter- and intra-gate correlations of spin and raising charge \eqref{eq:scorrinter} have an identical structure, as can be seen from the transition amplitudes in Appendix~\ref{sec:circuitaction}. Then, if we work \emph{in the frame of the front}, so that all distances below refer to distances away from the front location which is $t$ at infinite $q$, we can easily derive a recursion relation for the steady state values of $G_{ij}$ \eqref{eq:Cij} which takes the form:
\begin{equation}
G_{2j, 2k} = \frac{1}{3} (G_{2j, 2k-2} + G_{2j-2, 2k} + G_{2j-2, 2k-2}).
\end{equation}
Due to the even-odd structure of the circuit, the correlations involving odd sites or pairs of even and odd sites can all be derived from the correlations of the even sites alone. Solving this recursion relation by standard methods gives the solution 
\begin{equation}
G_{2j, 2k} = \frac{1}{3} \sum_{n = k}^{j+k} \left(\frac{1}{3} \right)^n {{n}\choose{ 2n - (j+k)}}{{2n - (j+k)}\choose{ n-j}}
\end{equation}
which, after taking care of even-odd effects reduces to \eqref{eq:deltars_exact} at $j=k$. 

\end{appendix}

\end{document}